\documentclass[12pt]{article}
\usepackage{graphicx,epsfig}
\usepackage{amsmath,amssymb}
\usepackage{array}
\usepackage{arydshln}
\usepackage{color}
\usepackage{cite}

\newcommand{\ord}[1]{\mathcal{O}\left({#1}\right)}

\title{\Large\bf\boldmath
RGE Behaviour of SUSY with a $U(2)^3$ symmetry
}
\date{}
\author{Gianluca~Blankenburg$^{a}$, Joel~Jones-P\'erez$^{b,c,d}$
\\[0.5 cm]
$^a${\em\normalsize Dipartimento di Fisica, Universit\`a di Roma Tre,} \\
{\em\normalsize  Via della Vasca Navale 84, I-00146 Roma, Italy} \\
$^b${\em\normalsize INFN, Laboratori Nazionali di Frascati, Via E.~Fermi 40, I-00044 Frascati, Italy} \\
$^c${\em\normalsize Departament de F\'{\i}sica Te\`orica and IFIC, Universitat de Val\`encia-CSIC,} \\
{\em\normalsize E-46100, Burjassot, Spain} \\
$^d${\em\normalsize CERN, Theory Division, 1211 Geneva 23, Switzerland} 
}

\begin{document}

\maketitle

\begin{abstract}
\noindent
The first LHC results seem to disfavour, from the point of view of naturalness, any constrained MSSM realization with universal conditions at the SUSY-breaking scale. A more motivated scenario is given by split-family SUSY, in which the first two generations of squarks are heavy, compatible with a $U(2)^3$ flavour symmetry. We consider this flavour symmetry to be broken at a very high scale and study the consequences at low energies through its RGE evolution. Initial conditions compatible with a split scenario are found, and the preservation of correlations from minimal $U(2)^3$ breaking are checked. The various chiral operators in $\Delta F=2$ processes are analyzed, and we show that, due to LHC gluino bounds, the $(LL)(RR)$ operators can not always be neglected. Finally, we also study a possible extension of the $U(2)^3$ model compatible with the lepton sector.
\end{abstract}

\section{Introduction}

In the past years, weak scale supersymmetry (SUSY) has possibly been the best motivated paradigm as an extension for the Standard Model (SM).

The most popular realization of SUSY has been the Minimal Supersymmetric Standard Model (MSSM), usually studied through its constrained form, the CMSSM. Based on models of gravity mediated SUSY-breaking, the CMSSM has only a few additional parameters, defined at a unification scale, $M_{\rm GUT}$. The parameters of the model would then be evolved down to a SUSY-decoupling scale, $M_{\rm SUSY}$, through the use of renormalization group equations (RGE). One of the main consequences of this approach is that the soft SUSY-breaking terms would acquire a flavour structure compatible with the Minimal Flavour Violation (MFV) ansatz~\cite{D'Ambrosio:2002ex}.

The current lack of SUSY signals at the LHC have forced the theoretical community to start stepping way from this simple realization of SUSY, and consider more complex scenarios that could still preserve all of its virtues, but at the same time avoid the current experimental constraints. One such scenario is that of hierarchical soft sfermion masses~\cite{Barbieri:2010pd, Giudice:2008uk, Cohen:1996vb,Dimopoulos:1995mi,Dine:1993np,Badziak:2012rf}, where the first two generations are heavy enough to avoid the current bounds, but the third generation is light enough to mitigate the naturalness problem, and still give us a chance to observe a signal in the near future.

Given such a scenario, one can question if a MFV flavour structure is as well motivated as in the CMSSM. In fact, it is difficult for MFV to provide such a split layout, especially for very hierarchical masses. Thus, it is of interest to study frameworks similar to MFV, but that can accomodate soft masses that are not degenerate. This motivation is further encouraged by the existence of a small tension between CP-violation (CPV) observables in $\Delta F=2$ observables in the $K$ and $B_d$ sectors, which MFV is unable to solve~\cite{Lunghi:2008aa,Buras:2008nn,Altmannshofer:2009ne,Buras:2009pj}. By stepping away from MFV, one would hope that the new framework would ameriolate this tension, and give us further insight on the flavour structure of the MSSM itself.

Such a framework was found in~\cite{Barbieri:2011ci}, based on a $U(2)^3$ flavour symmetry applied on the quark sector, broken in a minimal way by spurion fields\footnote{Examples of $U(2)^3$ being applied on non-SUSY scenarios can be found in~\cite{Feldmann:2008ja,Kagan:2009bn}.}. The framework successfully solved the flavour tension, predicting rather light third generation and gluino masses, as well as a somewhat large CPV phase on the $B_s$ sector. The phenomenological consequences in the quark sector were expanded in~\cite{Barbieri:2011fc,Barbieri:2012uh,Buras:2012sd}, and the neutrino and slepton sector were considered in~\cite{Blankenburg:2012nx}. A non-minimal realization of the framework was also studied in~\cite{Barbieri:2012bh}.

Nevertheless, all of these works considered the $U(2)^3$ symmetry to be directly applied at the electroweak scale, while the typical expectation is for this symmetry to be broken at a very high scale. This leads to several questions to be pondered. First of all, it is unclear if the running of the MSSM parameters would preserve the virtues following from the assumption of a minimal breaking of the $U(2)^3$ symmetry, analysed in the previous literature. Moreover, the type of initial conditions required to achieve the split scenario are not evident, especially after applying the LHC bounds on the gluino mass and trying to mitigate the naturalness problem as well as possible.

In this work, we attempt to answer these questions. To this end, in Section~\ref{sec:u2} we analyze the RGE behaviour of the minimal $U(3)^3$ framework defined in~\cite{Barbieri:2011ci}. We begin by analyzing a CMSSM-like parameter space in Section~\ref{sec:u2.spectrum}, and defining two Benchmark points of interest. We follow the evolution of the mixing in Section~\ref{sec:u2.mixing}, and evaluate how well the features of the symmetry are preserved in Section~\ref{sec:u2.structure}. Section~\ref{sec:u2.othoperator} is devoted to understanding the importance of each effective operator in the $\Delta F=2$ analysis, at the scale where SUSY decouples. Finally, in Section~\ref{sec:u2dev}, we consider a deviation from the $U(2)^3$ framework worked out in~\cite{Blankenburg:2012nx}, which was found necessary in order to reproduce the neutrino sector.

\section{$U(2)^3$ in the Quark Sector}
\label{sec:u2}

As mentioned in the introduction, in most works studying the $U(2)^3$ framework~\cite{Barbieri:2011ci, Barbieri:2011fc, Barbieri:2012uh, Buras:2012sd}, the flavour structures are taken directly at a low scale, of the order of the TeV. Here, on the contrary, we are interested in the possibility that both supersymmetry and the flavour symmetries are broken at a very high scale, that we take to be $M_{\rm GUT}\sim  10^{16}$ GeV.

We follow a three step procedure. In the first step, we define what our initial conditions shall be at $M_{\rm GUT}$. The fermion Yukawa couplings are determined from their electroweak values~\cite{Xing:2007fb, Flavour:CKM}, running them to the unification scale. For the sfermions, we assume the soft masses to follow the structures outlined in the Appendix, in particular, Eqs.~(\ref{md:LLGUT})-(\ref{mqLL:GUT}). We choose a common mass for all the squarks in the first two generations, $m_{heavy}$, and a common mass for the third generation, $m_{light}$. In general, we shall refer to the splitting $\rho=(m^2_{heavy}-m^2_{light})/m^2_{heavy}$ instead of $m_{light}$, such that $\rho=0$ is the totally degenerate case, while $\rho=1$ corresponds to maximal splitting. The A-terms follow a flavour structure similar to that of the Yukawas, but the different $\ord{1}$s lead to a non-diagonal structure, shown in Eq.~(\ref{trilinear:GUT}). We assume the A-terms to be connected with the first two generations masses, so we take a universal $A_0 \sim m_{heavy}$. This is done with the intention of enhancing stop mixing, such that an appropriate value for the Higgs mass is obtained. Moreover, we typically expect the leading soft scale, and so $A_0$, to be related with $m_{heavy}$ and that $m_{light}$ can be obtained through some subleading mechanism (for an example, see~\cite{Barbieri:2011ci}).

Regarding the other soft parameters, we use a common value for the Higgs soft masses $m_{H_u}=m_{H_d}=m_0$, that can be different from both the light and heavy sfermion masses. We also consider a unified gaugino mass $M_{1/2}$, and, for definiteness, we fix $\tan{\beta}=10$. Thus, in our scan, the variable parameters shall be:
\begin{equation}
 M_{1/2},\,m_{heavy},\,\rho,\,m_0,\,A_0,\, x_i,\, \gamma_i~,
\end{equation}
where $x_i$ and $\gamma_i$ represent the $\ord{1}$ parameters and phases shown in the Appendix. Notice that $A_0$ shall actually be the product of $m_{heavy}$ and an $\ord{1}$ parameter.

On the second step, we run all parameters down to a common decoupling scale $M_{\rm SUSY}\sim1$~TeV, following 2-loop MSSM RGEs~\cite{Martin:1993zk}. As we are interested entirely in the RGE effects, we do not consider threshold corrections. At this scale, we calculate the sfermion soft masses and mixings.


Once we are at the low scale, we proceed with the third step, which is to ask several requirements to be satisfied. First, we ask the absence of color/charge-breaking minima. In fact, the third generations masses can acquire tachyonic values due to negative contributions from the running, proportional to $m_0$ at one loop and to $m_{heavy}$ at two loops~\cite{ArkaniHamed:1997ab}. On the contrary, $M_{1/2}$ induces a positive contributions to the running, pushing the sfermion mass towards positive values, while the influence of the A-terms is weak. Thus, a balance between all contributions shall be required, such that no sfermion masses become tachyonic. As we shall see, the tachyon bound can forbid some scenarios with large splitting, $\rho\sim1$\footnote{Notice that, as we are taking a universal decoupling scale, the real edge of the tachyon bound is probably less stringent than the one shown. As shown in~\cite{Tamarit:2012ie}, the early decoupling of the heavy first generations can prevent some cases from becoming tachyonic.}.


We also ask for correct radiative electroweak symmetry breaking (EWSB), ie.\ we demand that at the low scale the $\mu$ parameter has a value such that the tapdole terms of the scalar potential vanish. For this to be satisfied, it is usually sufficient to have $m_{H_u}^2$ acquiring a negative value due to the running. Thus, a large $m_0$ shall be disfavoured, as it will imply that the initial value of $m_{H_u}$ shall be too large in order to be driven negative from the running. Moreover, a small $M_{1/2}$ is also indirectly disfavoured. This is due to the negative influence of the stop masses on $m_{H_u}$. Larger values of $M_{1/2}$ shall give a larger positive gluino contribution to the stop masses, which in turn shall provide a larger negative contribution to $m_{H_u}^2$.

In addition, we require LHC bounds to be satisfied. In particular, we demand the light Higgs mass $m_h$ to be compatible with the latest ATLAS and CMS measurements, that is, we take $m_h^{exp}=125.3 \pm 0.6$ GeV~\cite{:2012gu, :2012gk}. We calculate both $m_h$ and its theoretical error using FeynHiggs~\cite{Heinemeyer:1998yj, Heinemeyer:1998np, Degrassi:2002fi, Frank:2006yh}, bounding the latter to be no larger than 3 GeV. We then ask $m_h$ to be within the $1\sigma$ range, which in principle can allow masses as small as 123 GeV. We also check that all direct SUSY bounds are satisfied (in particular, $m_{\tilde{g}}>1$ TeV is the most relevant limit).

Moreover, an interesting feature of the $U(2)^3$ model, in the flavour sector, is that it can improve the CKM fit with tiny and correlated new contributions to $\epsilon_K$ and $S_{\psi K_S}$~\cite{Barbieri:2011ci, Buras:2012sd}. The size of these gluino-mediated effects depends on the function $F_0$, defined as:
\begin{eqnarray}
F_0 &=& \frac{2}{3} \left(\frac{g_s}{g} \right)^4 \frac{ m_W^2 }{ m^2_{Q_3} } \frac{1}{S_0(x_t)}
\left[ f_{0}\left(\frac{ m^2_{\tilde g} }{ m_{Q_3}^2 }\right) + \ord{\frac{m_{Q_l}^2}{m_{Q_h}^2}}\right]~, 
\label{eq:F0} \\
f_{0}(x) & = & \frac{11 + 8 x -19x^2 +26 x\log(x)+4x^2\log(x)}{3(1-x)^3}~, 
\end{eqnarray}
with $S_0(x_t)$ being the typical one-loop function of the SM to $\Delta F=2$ processes (for example, see~\cite{Buchalla:1995vs}). In the updated fit of~\cite{Straub:2011fs}, it was shown that, after the inclusion of LHCb data, the $U(2)^3$ contributions could be of the correct size to solve the flavour tension if $0.01<F_0<0.14$, and if the mixing was above a certain value\footnote{This assumes $|V_{ub}|=(3.97\pm0.45)\times10^{-3}$.}. As can be expected, the requirement on $F_0$ can be satisfed only if the $\tilde{g}$ and $\tilde{b}_L$ are not too heavy. We will mark this region with a special line in the following plots.

Given all these constraints, we will concentrate on the regions with a soft spectrum at $M_{\rm SUSY}$ as natural as possible. In particular, as shown in the literature~\cite{Papucci:2011wy, Baer:2012vr, Dimopoulos:1995mi}, a natural supersymmetric theory requires that $\mu$ and the third generation masses to be light, and that the gluino must not be too heavy. Note that this represents a tension with the value of the parameters required to obtain a Higgs mass heavier than 120 GeV. Thus, we shall place ourselves in a middle ground, searching for values of $\mu<1$~TeV, and at least one stop with mass $m_{\tilde t_1}< 1$~TeV.

We are also interested in understanding the type of splitting one obtains after the running. We shall be presenting our results in terms of:
\begin{equation}
\rho^{\rm low}_{\tilde t,\tilde b}=\frac{\langle m^2_{sq}\rangle-\langle m^2_{\tilde t,\tilde b}\rangle}{\langle m^2_{sq}\rangle}
\end{equation}
where $\langle m^2_{\tilde t,\tilde b}\rangle$ is the average mass squared for the stops or sbottoms, and $\langle m^2_{sq}\rangle$ is the average mass squared of the respective first two generations.

\subsection{Spectrum}
\label{sec:u2.spectrum}

In our study we focus on two different scenarios. First, we shall take $m_{heavy}=3$~TeV, very close to the experimental limit, such that it might be feasible to observe some signals from the first two generation squarks in the near future. On the second scenario, we use $m_{heavy}=8$~TeV, and see what consequences this has on the spectrum.

For each value of $m_{heavy}$, we need to evaluate the interplay between the values of $M_{1/2}$, $\rho$, $m_0$ and $A_0$ required to satisfy the bounds mentioned previously. We shall explain such an interplay around the following two Benchmarks:
\begin{description}
 \item[Benchmark 1:] $M_{1/2}=500$~GeV, $\rho=0.5$, $m_{heavy}=3$~TeV, $m_0=2.8$~TeV, $A_0=-m_{heavy}$~,

 \item[Benchmark 2:] $M_{1/2}=1.1$~TeV, $\rho=0.97$, $m_{heavy}=8$~TeV, $m_0=2.5$~TeV, $A_0=-0.25m_{heavy}$~,
 \end{description}
which, as we shall see, satisfy all our requirements. For each Benchmark, the value of $A_0$ has been chosen in order to maximize stop mixing without generating tachyons, such that the appropriate Higgs mass is reproduced. The choice for the other parameters shall be made clear when examining the surrounding parameter space.

\begin{figure}[tbp]
\begin{center}
$\mu$~(GeV) \\
\includegraphics[width=0.32\textwidth]{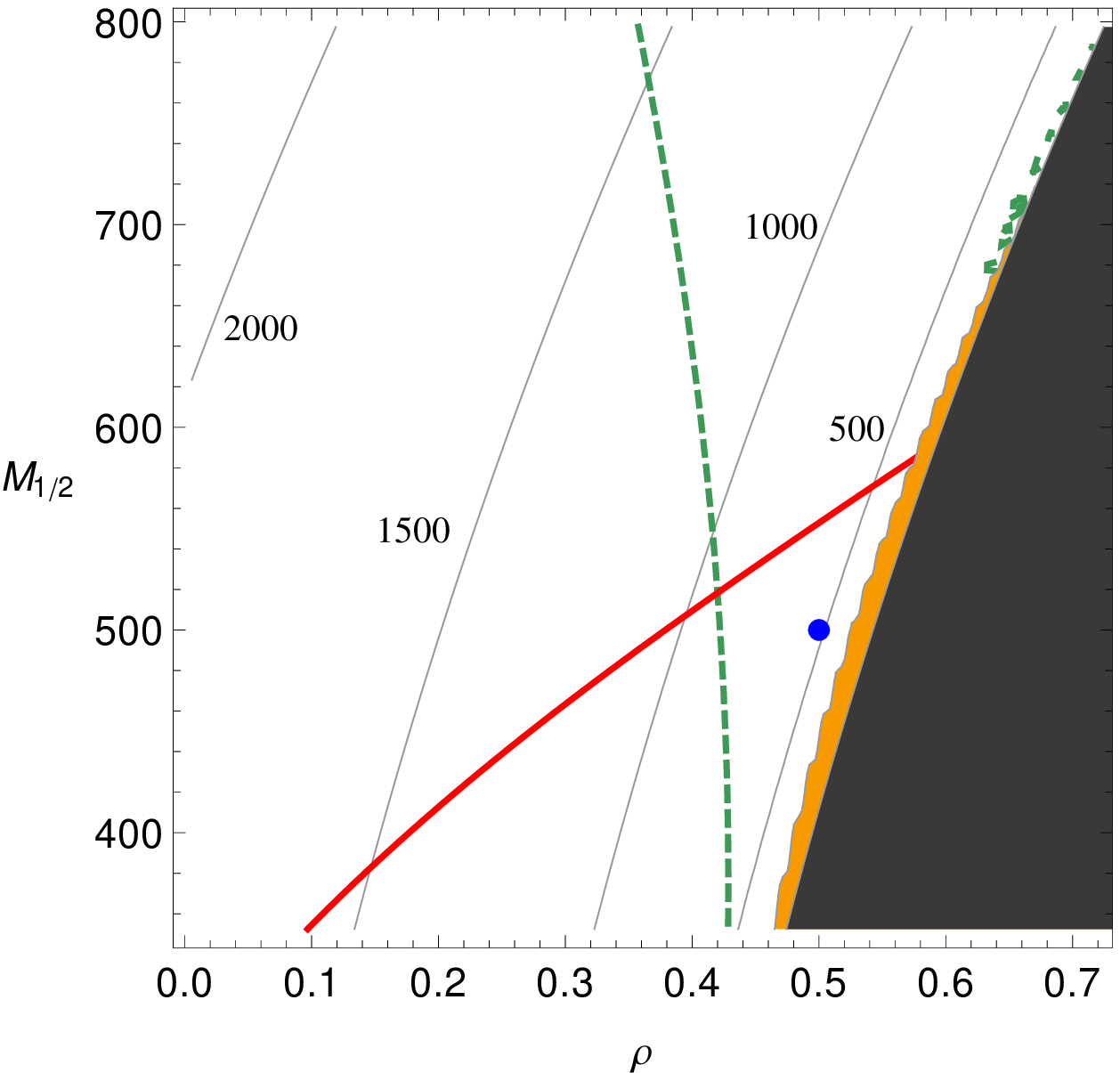}
\includegraphics[width=0.32\textwidth]{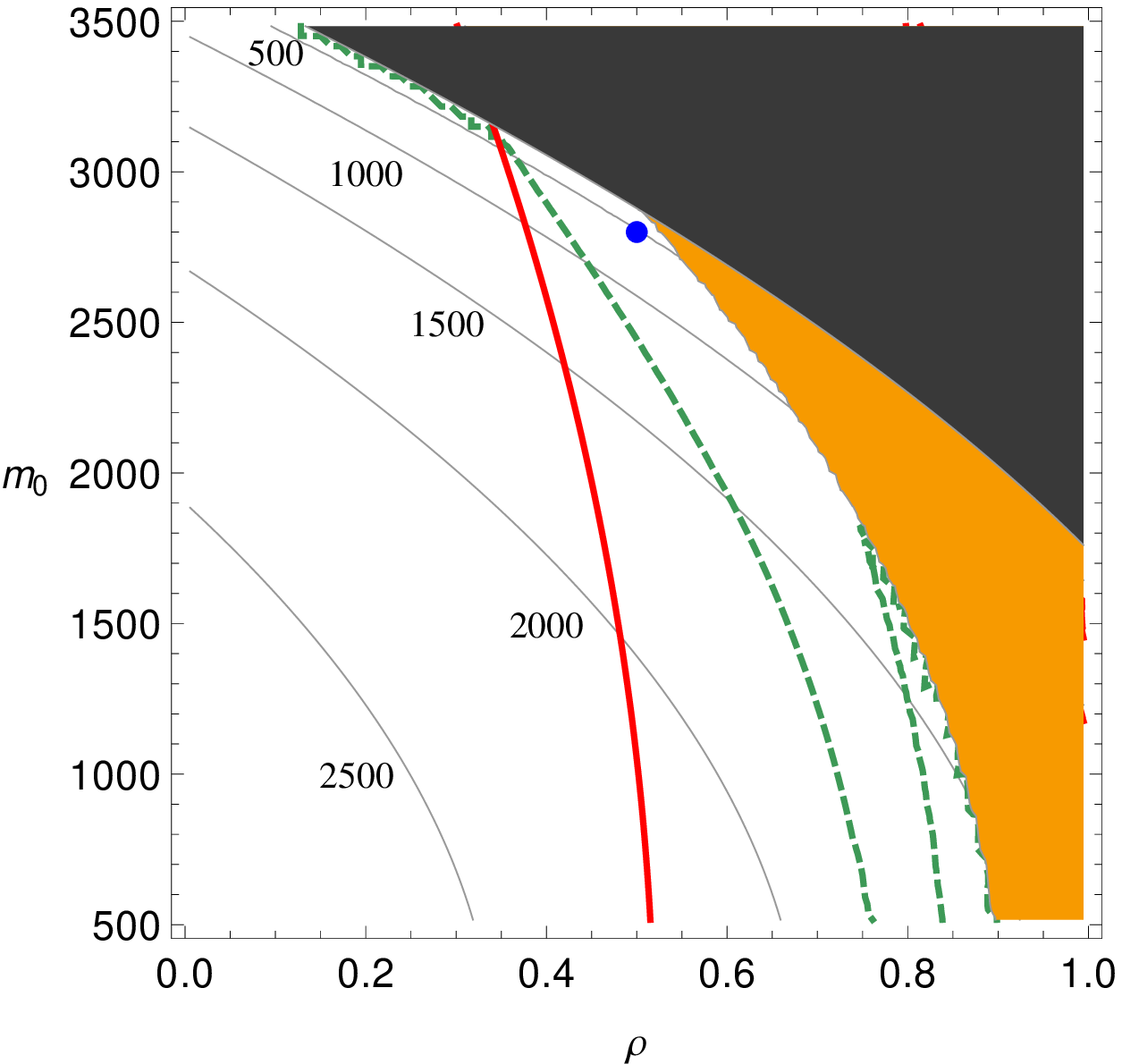}
\includegraphics[width=0.32\textwidth]{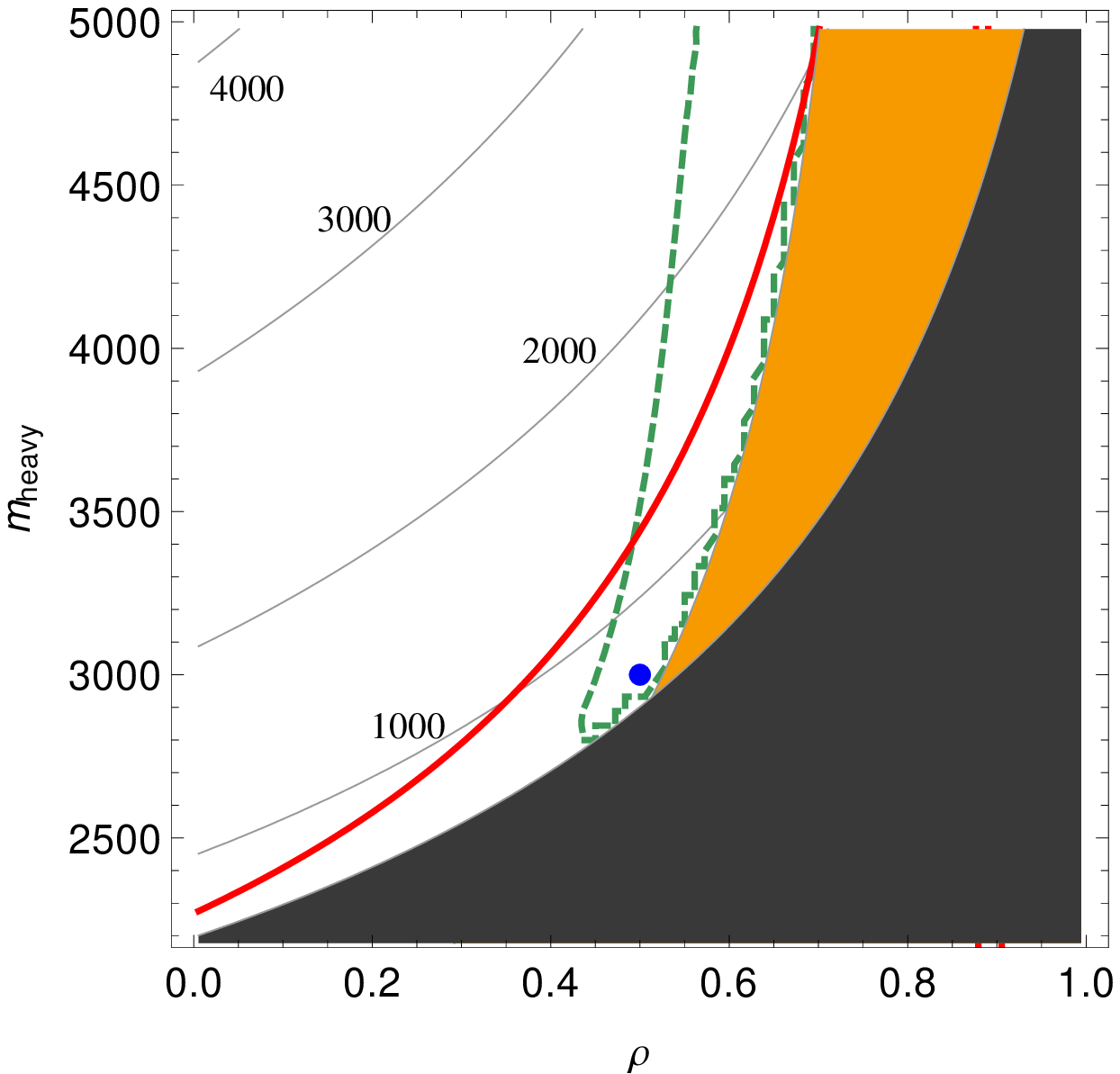} \\
$\rho^{\rm low}_{\tilde t}$ \\
\includegraphics[width=0.32\textwidth]{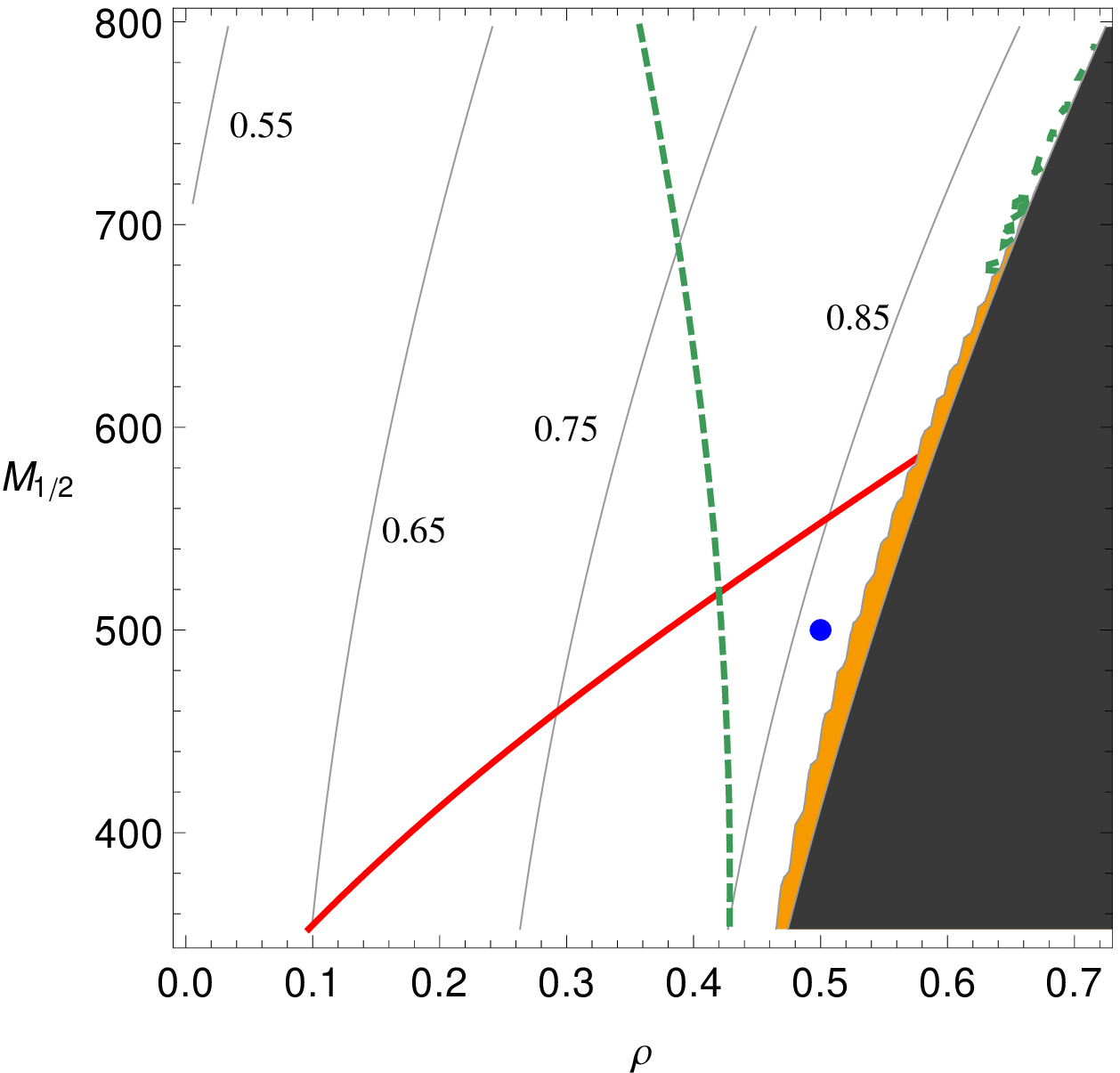}
\includegraphics[width=0.32\textwidth]{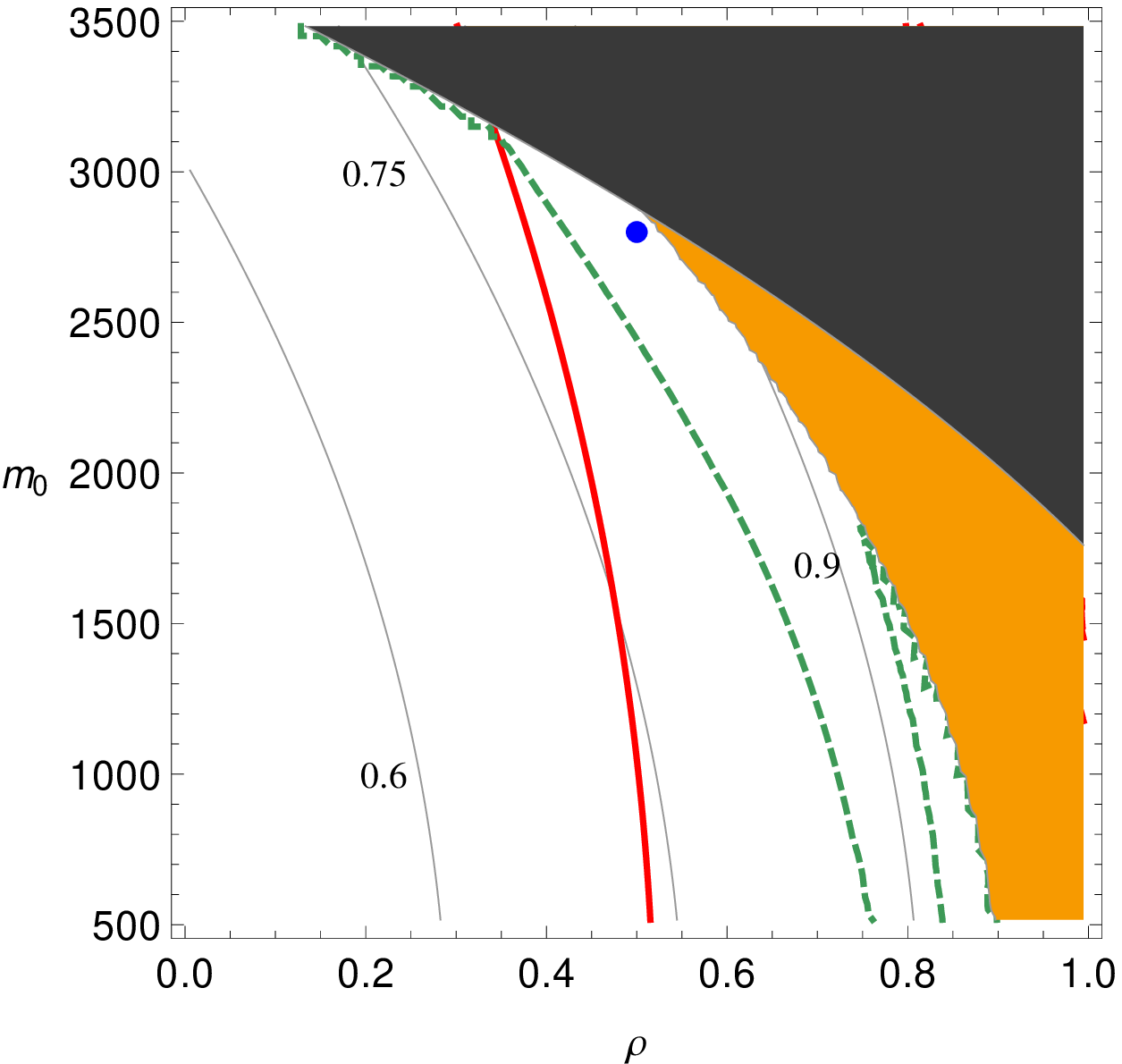}
\includegraphics[width=0.32\textwidth]{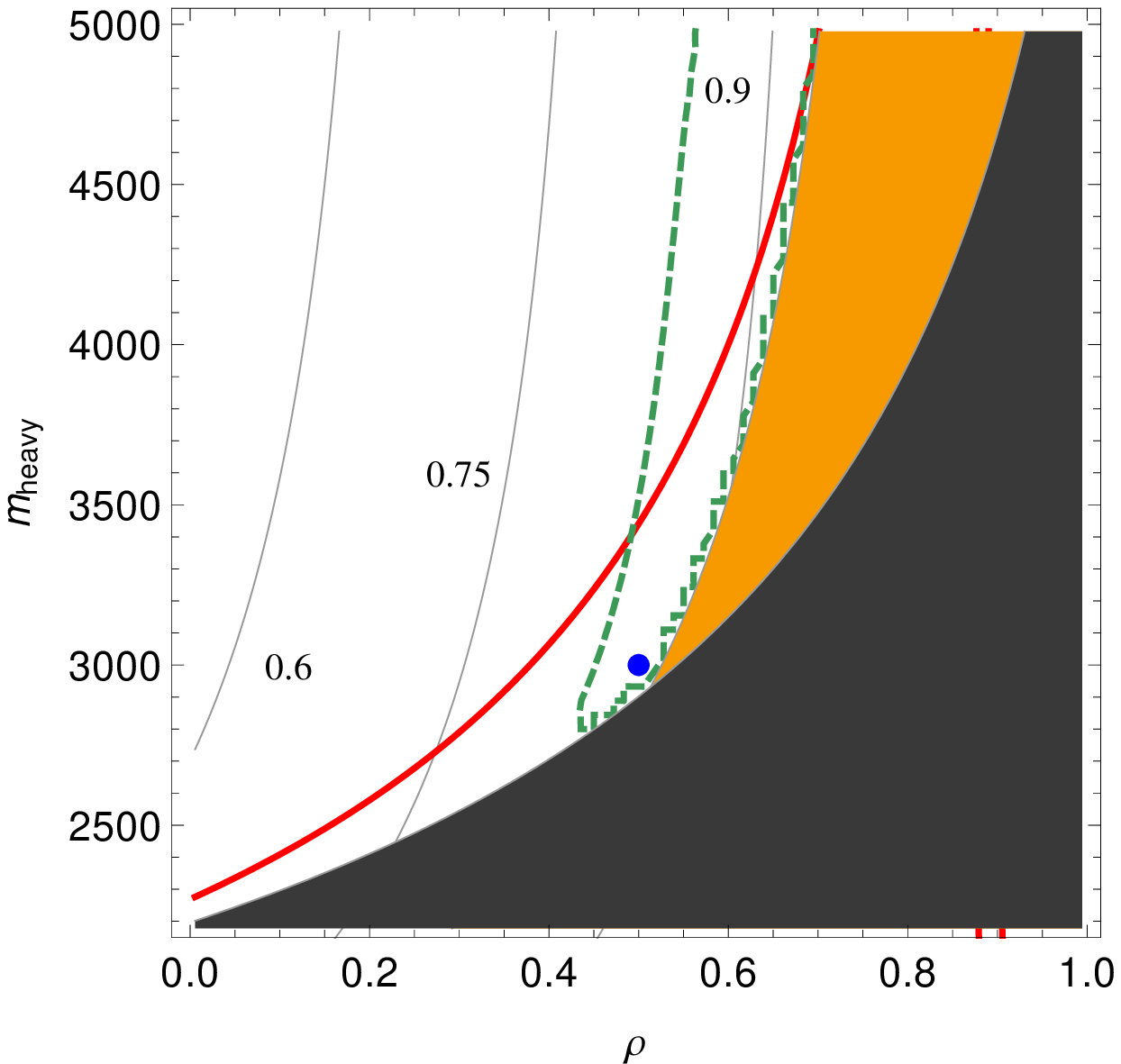} \\
$\rho^{\rm low}_{\tilde b}$ \\
\includegraphics[width=0.32\textwidth]{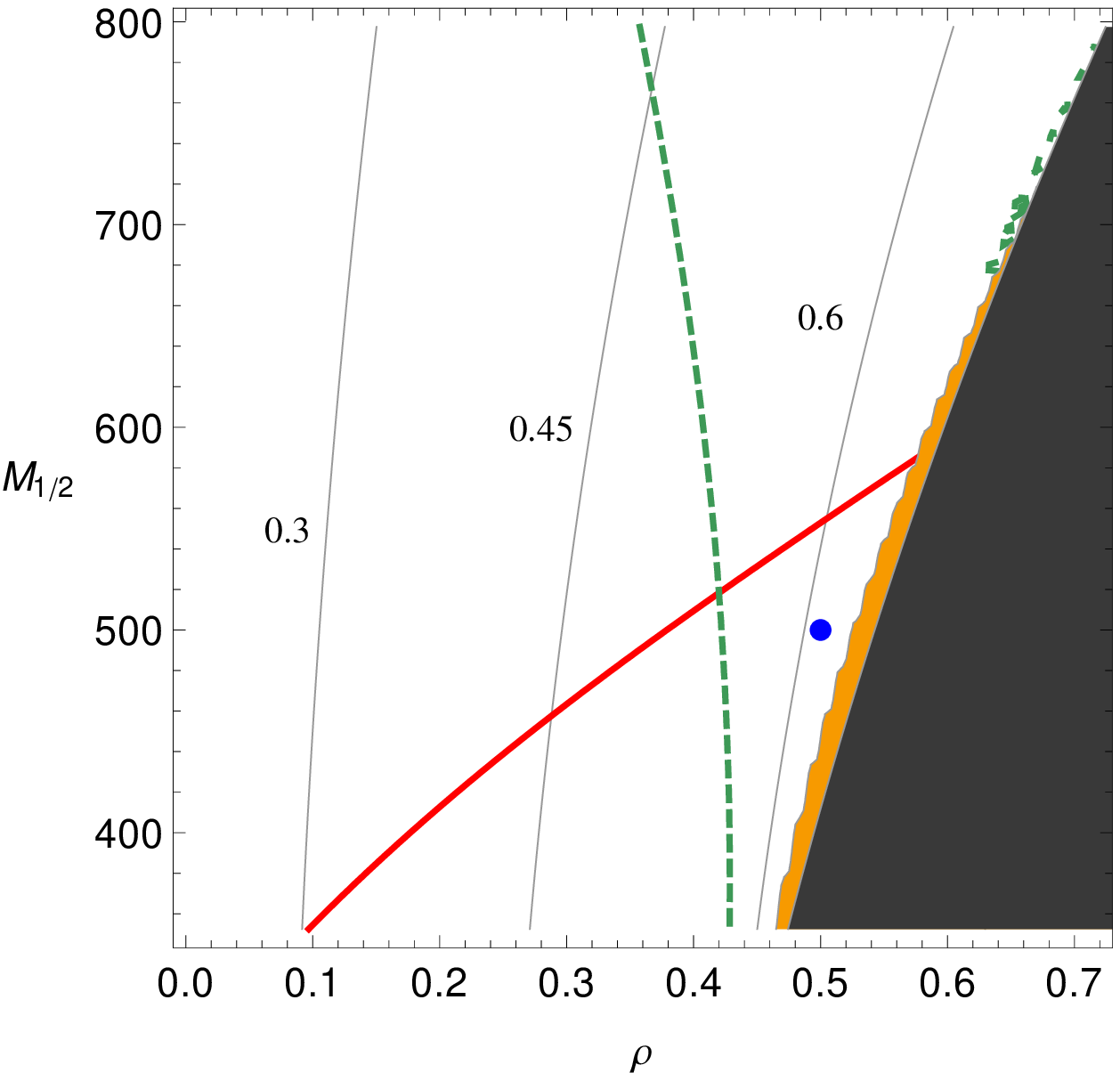}
\includegraphics[width=0.32\textwidth]{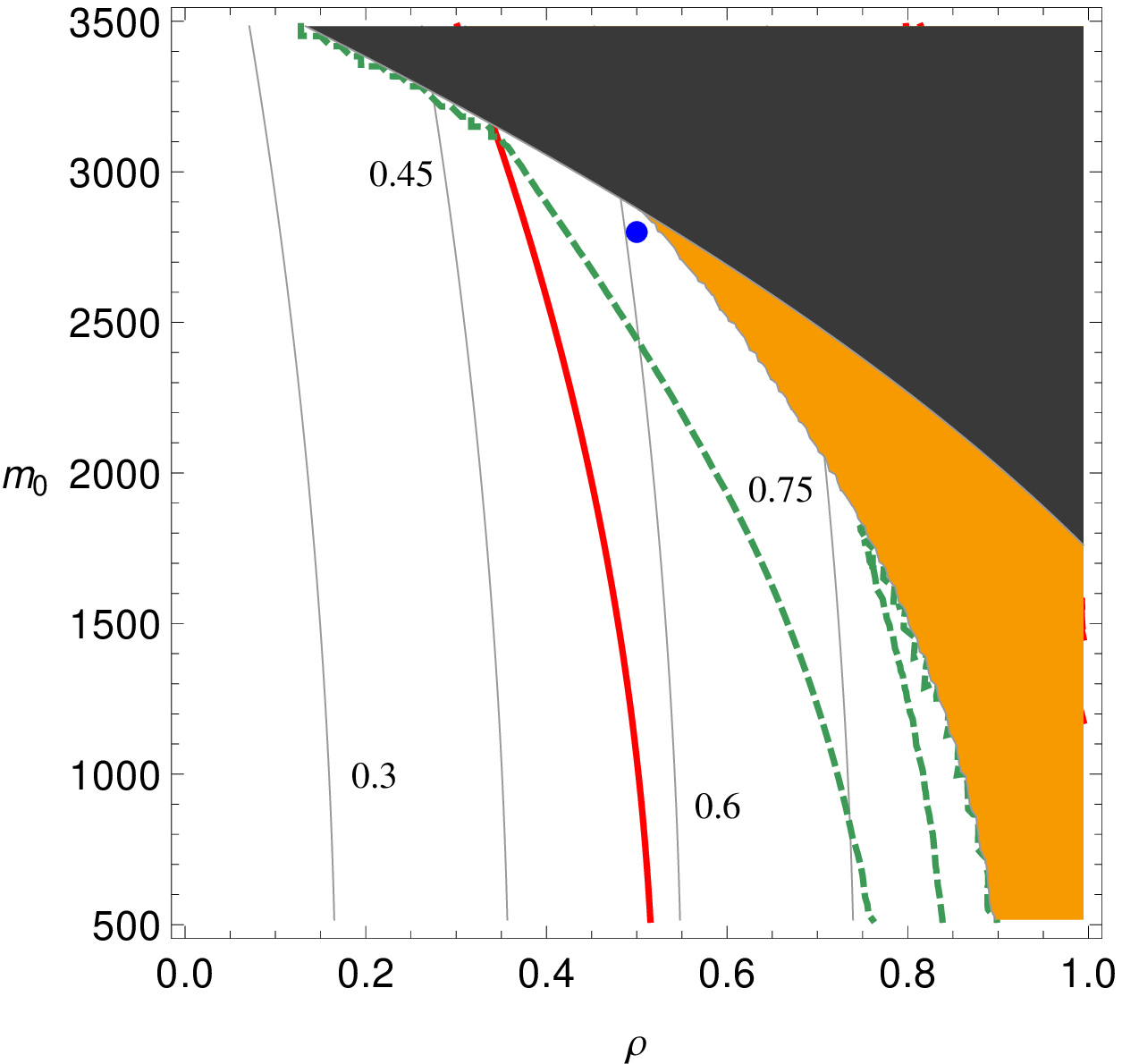}
\includegraphics[width=0.32\textwidth]{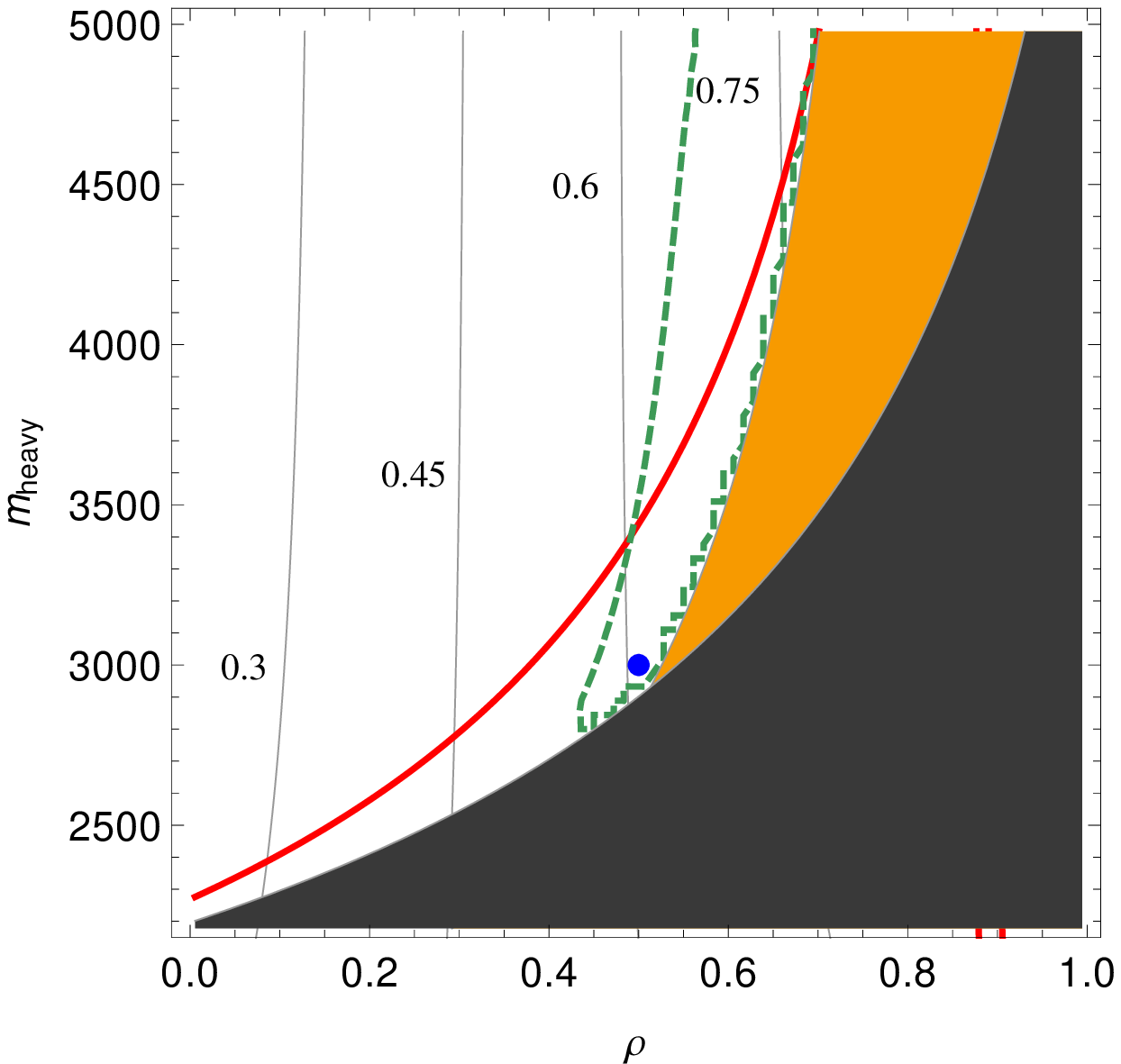}
\end{center}
\caption{Parameter space around Benchmark 1. We show contours for $\mu$, $\rho^{\rm low}_{\tilde t}$ and $\rho^{\rm low}_{\tilde b}$ on the top, centre and bottom, respectively. The dark regions correspond to no EWSB, while the orange regions have a tachyonic stop. The green, dashed lines delimitate the regions within 1 sigma of the Higgs mass, and the red, solid curve indicates the regions below which the flavour tension could be solved. The blue dot represents Benchmark 1, which satisfies all constraints.} 
\label{fig:benchmark1}
\end{figure}

\begin{figure}[tbp]
\begin{center}
$\mu$~(GeV) \\
\includegraphics[width=0.32\textwidth]{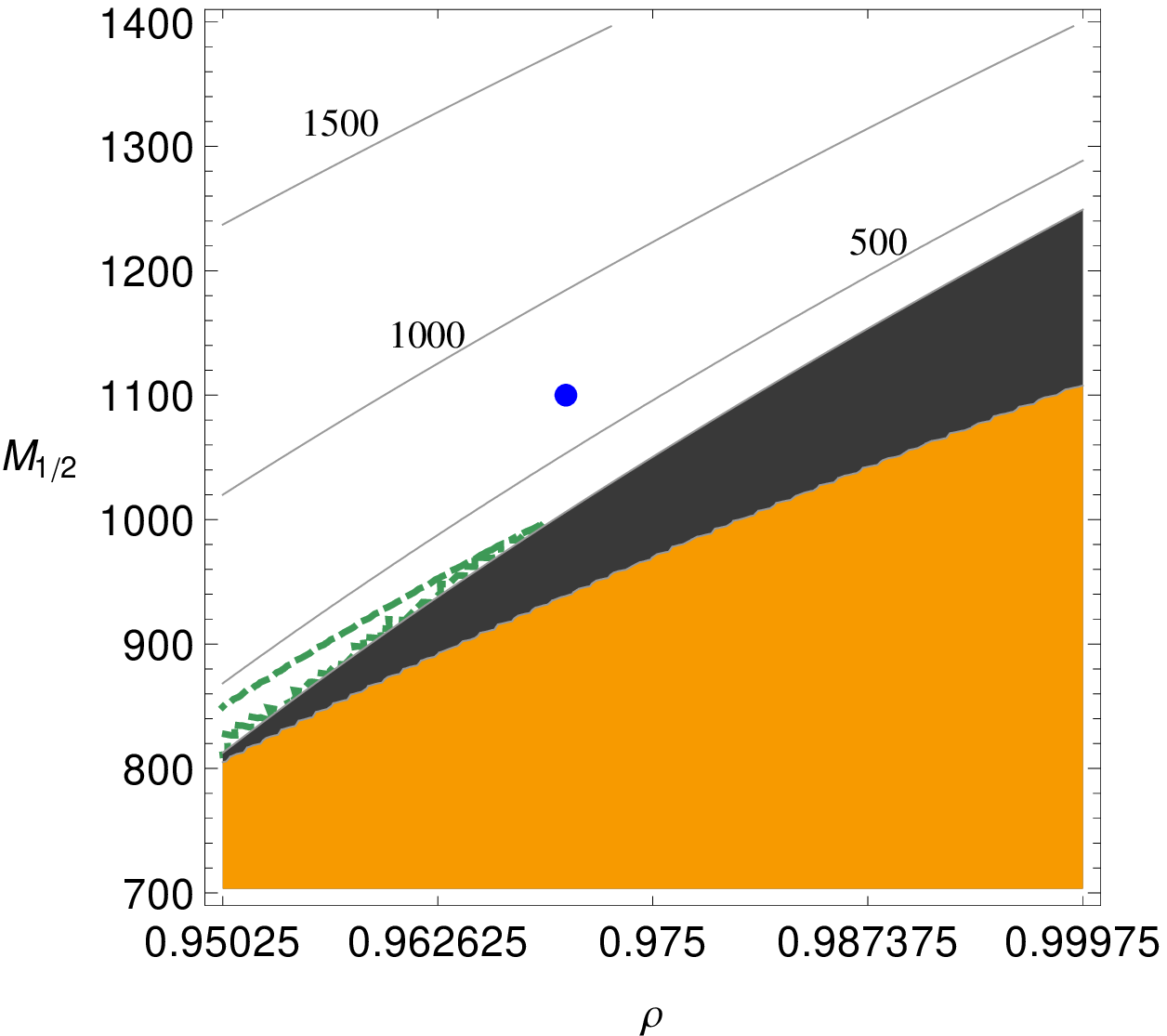}
\includegraphics[width=0.32\textwidth]{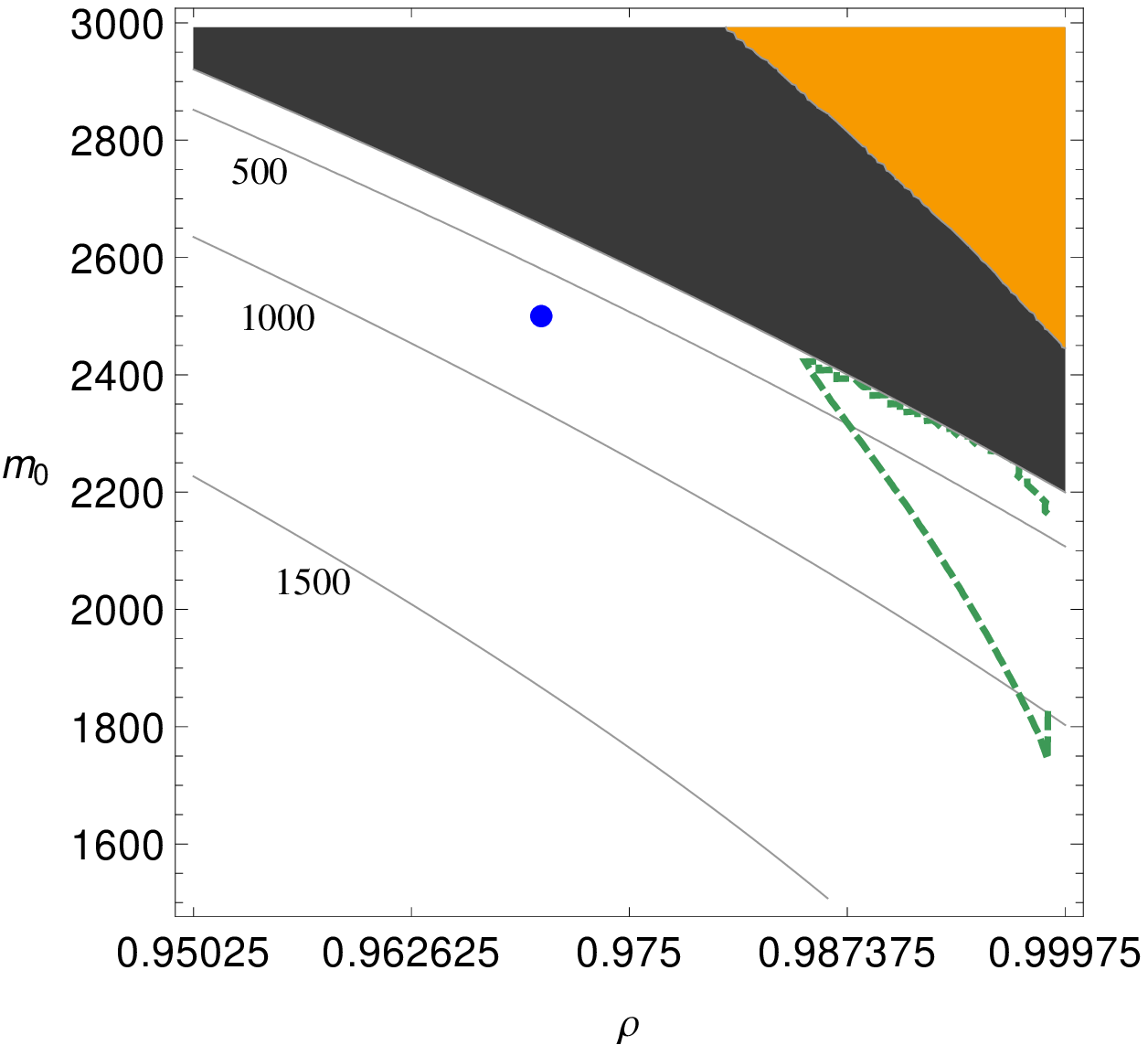}
\includegraphics[width=0.32\textwidth]{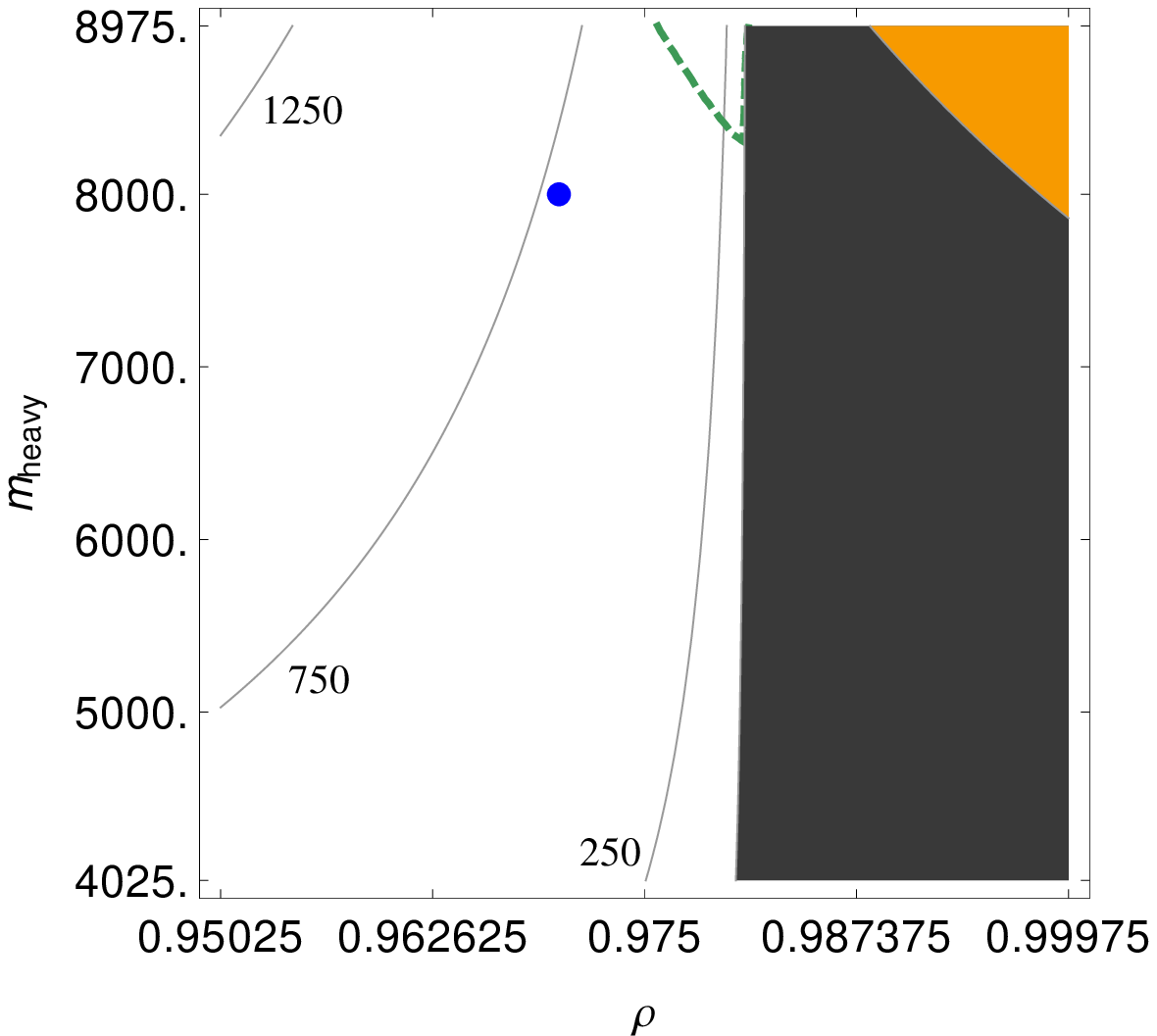} \\
$\rho^{\rm low}_{\tilde t}$ \\
\includegraphics[width=0.32\textwidth]{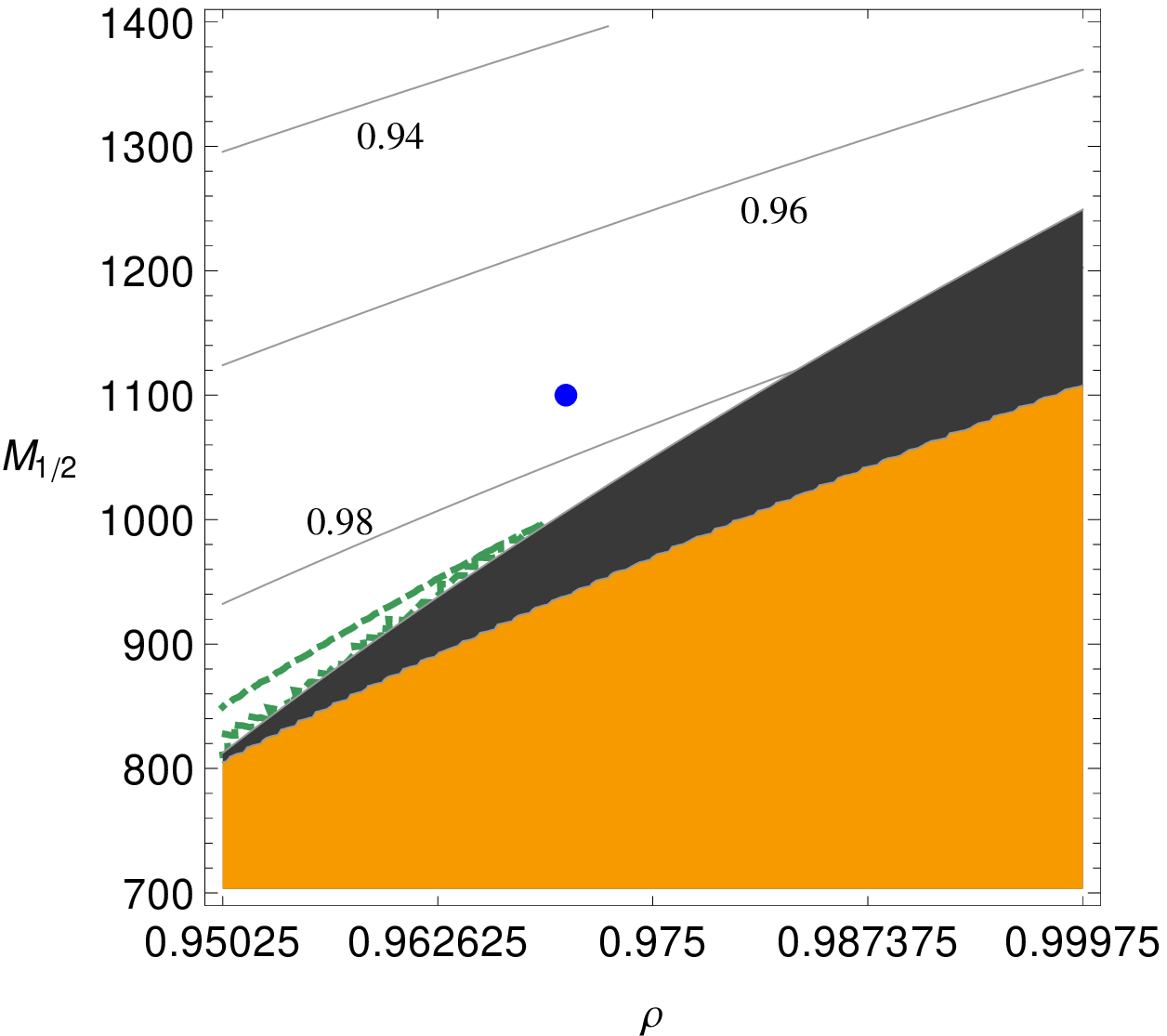}
\includegraphics[width=0.32\textwidth]{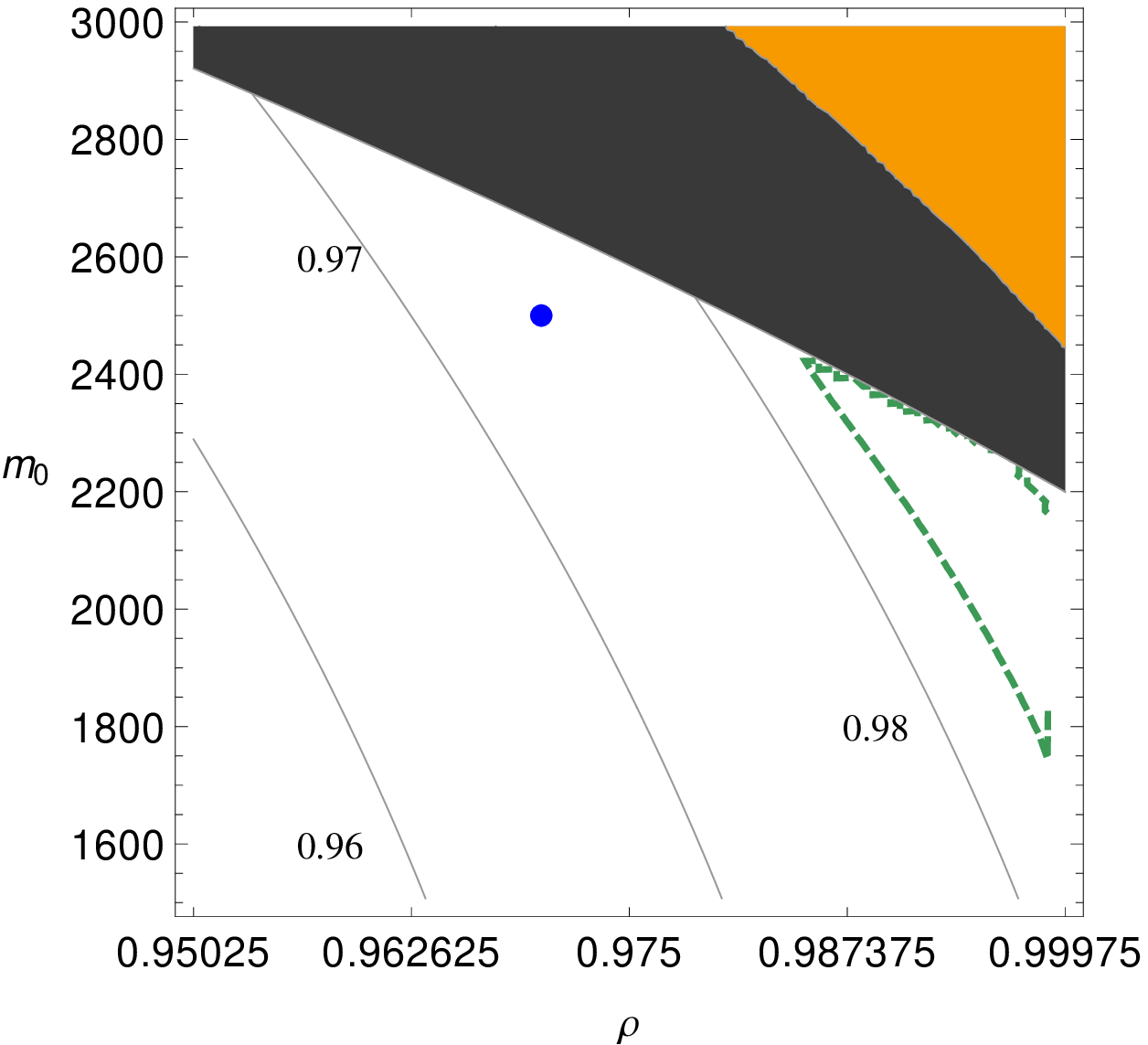}
\includegraphics[width=0.32\textwidth]{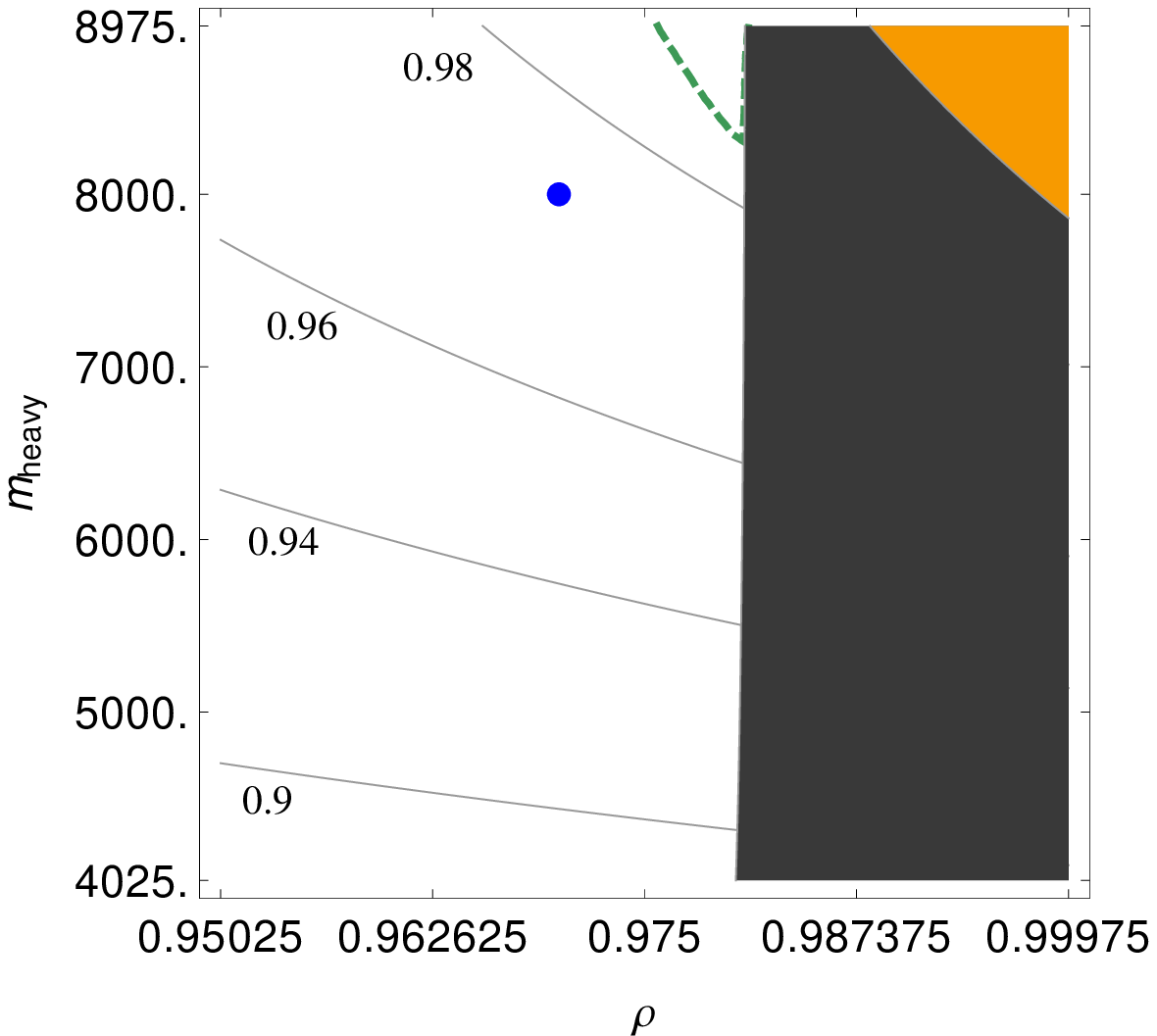} \\
$\rho^{\rm low}_{\tilde b}$ \\
\includegraphics[width=0.32\textwidth]{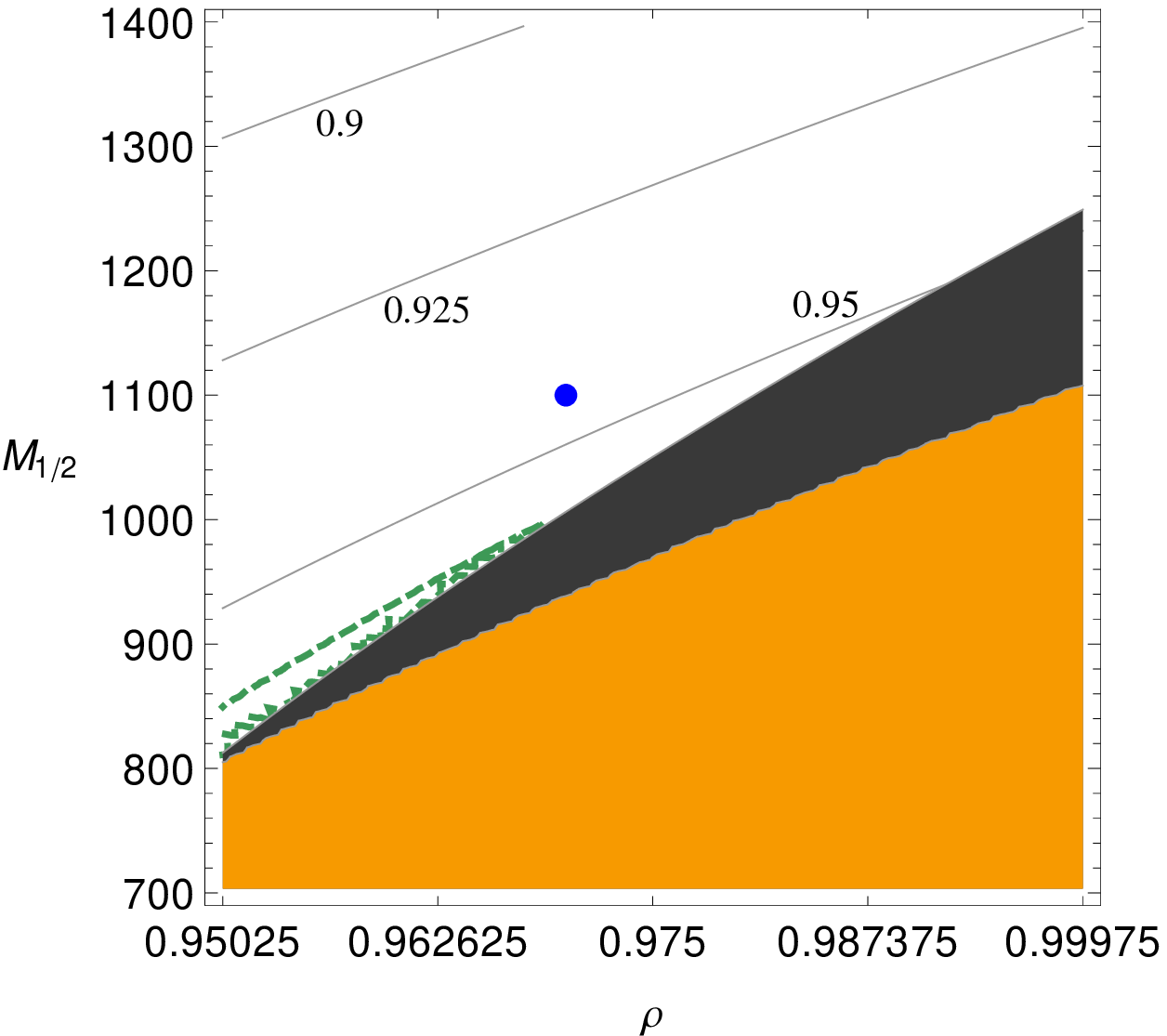}
\includegraphics[width=0.32\textwidth]{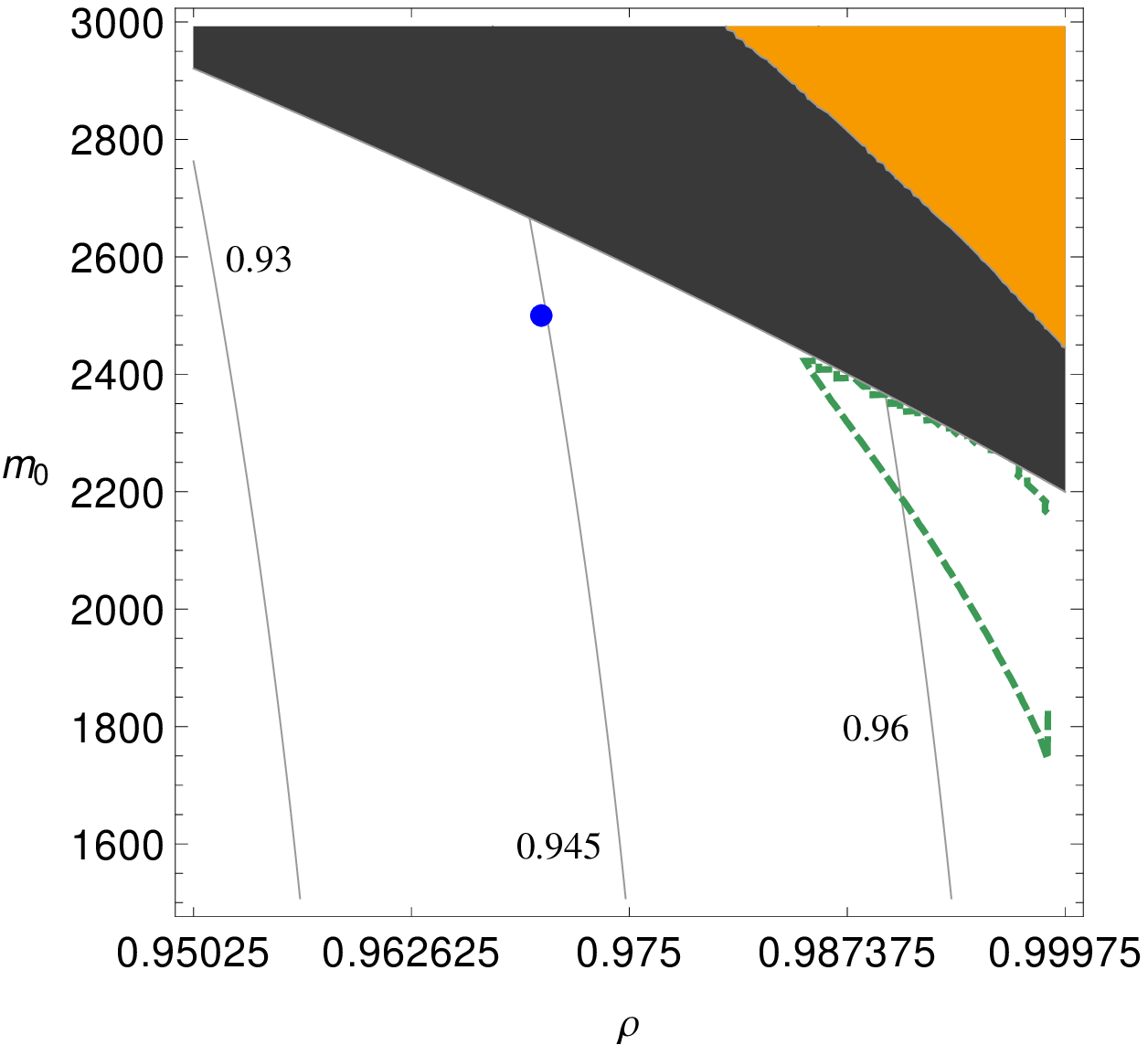}
\includegraphics[width=0.32\textwidth]{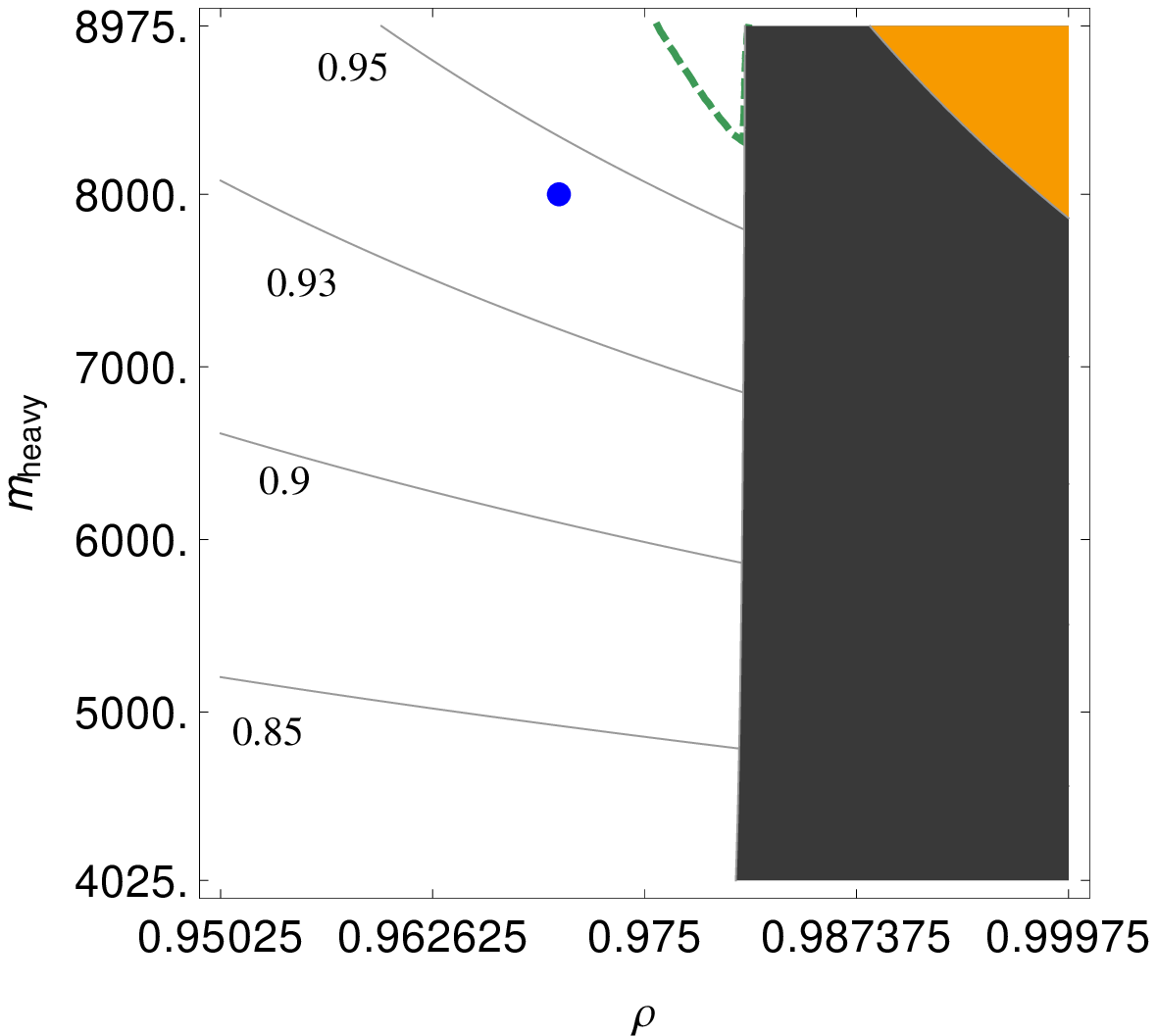}
\end{center}
\caption{Same as Figure~\ref{fig:benchmark1}, but for Benchmark 2. Notice that the evaluated parameter space never satisfies the $F_0$ constraint. The blue dot represents Benchmark 2, which satisfies all other constraints.} 
\label{fig:benchmark2}
\end{figure}

In Figures~\ref{fig:benchmark1} and~\ref{fig:benchmark2} we show contours of $\mu$, $\rho^{\rm low}_{\tilde t}$ and $\rho^{\rm low}_{\tilde b}$ on the upper, centre and lower rows, and show the interplay between $\rho$ and $M_{1/2}$, $m_0$ and $m_{heavy}$ on the left, centre and right columns, respectively. The dark regions correspond to points where EWSB is not achieved, and the orange regions have at least one tachyonic stop. We plot 1$\sigma$ bounds on the Higgs mass as a dashed, green line, and show the region satisfying the $F_0$ constraint with a solid, red line.

Let us focus on the parameter space around Benchmark 1, which is shown in Figure~\ref{fig:benchmark1}. Here, we find a strong upper bound on $\rho$ due to tachyons and EWSB. To avoid this bound, one can either increase $M_{1/2}$, decrease $m_0$ or increase $m_{heavy}$. However, as the Higgs and the $F_0$ bounds act in opposite directions, the possible variations are strongly limited. Increasing either $M_{1/2}$ or $m_{heavy}$ shall improve the Higgs mass, but at the same time will worsen the value of $F_0$. This situation is even further constrained by the naturalness bound on $\mu<1$~TeV, which fixes a lower bound on $m_0$. Similarly, the Higgs and $F_0$ constraints do not favour lower values of $\rho$.

Such constraints lead to values of $\mu$ around 500 GeV, and very specific splittings. In the stop sector, we find $\rho^{\rm low}_{\tilde t}\sim0.85$, which leads to an average stop mass of about $1.2$~TeV. Nevertheless, as the stop mixing is large, we find the mass of the lightest stop to be lower than 500~GeV. On the other hand, in the sbottom sector, we have $\rho^{\rm low}_{\tilde b}\sim0.6$, leading to an average sbottom mass of $1.9$~TeV. Notice that this setup involves a very mild splitting at the GUT scale, but can lead to a larger splitting in the stop sector. This is actually favoured by the neutrino sector, which was studied in a $U(2)^5$ framework generated from the breaking of $U(3)^5$~\cite{Blankenburg:2012nx}. Nevertheless, the splitting in the sbottom (and stau) sectors remains somewhat mild.

Let us turn now to Benchmark 2, shown in Figure~\ref{fig:benchmark2}. The motivation of this Benchmark is to study a scenario with a much stronger splitting than in Benchmark 1. However, the tachyon, Higgs and $\mu$ bounds force the value of $M_{1/2}$ to be too large to satisfy the $F_0$ constraint, having values about one order of magnitude lower than what is preferred from the fit in~\cite{Straub:2011fs}. Still, we consider Benchmark 2 a useful comparison, which might become of interest if a stop signal is observed in the upcoming data, with no corresponding gluino nor squark signal.

From Figure~\ref{fig:benchmark2}, we find that the splitting in the stop sector remains somewhat invariant. This means that the positive RGE contribution to $m_{\tilde t}$ from $M_{1/2}$ cancels the large, negative RGE contribution from the $y_t^2$ Yukawa and the two-loop contribution from $m_{heavy}$. However, in the sbottom sector, the $y_b^2$ contribution is not as large as that for the stop sector, so the splitting is somewhat reduced. Still, we have and average stop mass of about $1.4$~TeV (albeit with large mixing), an average sbottom mass close to 2~TeV, and $\mu=600$ GeV.

One must admit that such a heavy spectrum is less natural than that in Benchmark 1. Even though the $\mu$ parameter and the lightest stop mass are light enough, the gluino and the second stop are much heavier. Nevertheless, as this scenario reproduces the Higgs mass much easily that Benchmark 1, we still find this scenario attractive.

\subsection{Mixing}
\label{sec:u2.mixing}

Having found points in the parameter space leading to a split squark spectrum, we now turn to the question of what is the behaviour of the mixing after the running. This is crucial in order to understand if the results given in \cite{Barbieri:2011ci, Barbieri:2011fc} are modified if we take the flavour structures at $M_{\rm GUT}$ and then evolve them to the low scale.

When evaluating the method to track the RGE evolution of the mixing, one finds several choices. First, it is possible to study the variation of the off-diagonals through the mass-insertion approximation (MIA). However, as this framework provides a non-degenerate spectrum, it is unclear if the MIA is appropriate. Second, one could track the evolution of the $\ord{1}$ constants shown in Eqs.~(\ref{md:LLGUT})-(\ref{mqLL:GUT}), fitting the low-energy matrices into a $U(2)^3$-like structure. We find this procedure valid, but not particularly transparent nor informative. The third option is to build objects directly related to the physical observables, such that the evolution can be connected with the main results in~\cite{Barbieri:2011ci,Barbieri:2011fc}. As one can relate these objects with the framework parameters, we choose this approach.

As the communication between the first two and the third generations is due to a $U(2)_Q$ doublet, one would expect that the main deviation from MFV to be found in $m^2_{\tilde Q}$. This means that, if we concentrate in $\Delta F=2$ processes, the main supersymmetric contribution would come from $(LL)^2$ operators. Thus, we shall concentrate here on the evolution of the mixing participating in the latter operators, leaving the rest to be considered separately in Section~\ref{sec:u2.othoperator}.

From~\cite{Barbieri:2011ci}, we find that the gluino mediated contributions to $(LL)^2$ operators depend on the combination:
\begin{equation}
\lambda^{(a)}_{i\not = j} = (W^d_L)_{ia} (W^d_L)^*_{ja}~,
\end{equation}
where $W^d_L$ is the diagonalization matrix of $m_{\tilde{Q}}^2$ in the basis of diagonal down quarks. In particular
the supersymmetric contributions to $K$, $B_{d}$ and $B_{s}$ physics (in the limit of $\rho\to1$) are given respectively by
\begin{align}
\label{eq:lambda}
\lambda^{(3)}_{12}= s_L^2 \kappa^* c_d~, & & \lambda^{(3)}_{13}= -s_L \kappa^* e^{i\gamma_L}~, 
& & \lambda^{(3)}_{23}=-c_d s_L e^{i\gamma_L}~,
\end{align}
where $s_L=x_L\,\epsilon$ and $\kappa\approx c_d V_{td}/V_{ts}$, with the remaining parameters defined in the Appendix. We see that the only parameters not fixed by the CKM matrix are $s_L$ and $\gamma_L$, so the three objects are expected to be correlated. In Section~\ref{sec:u2.structure} we shall analyze how these correlations behave under the RGE evolution, so for now it suffices to consider only the evolution of one of these objects. We shall choose $\lambda^{(3)}_{23}$.

\begin{figure}[tbp]
\begin{center}
\includegraphics[width=0.45\textwidth]{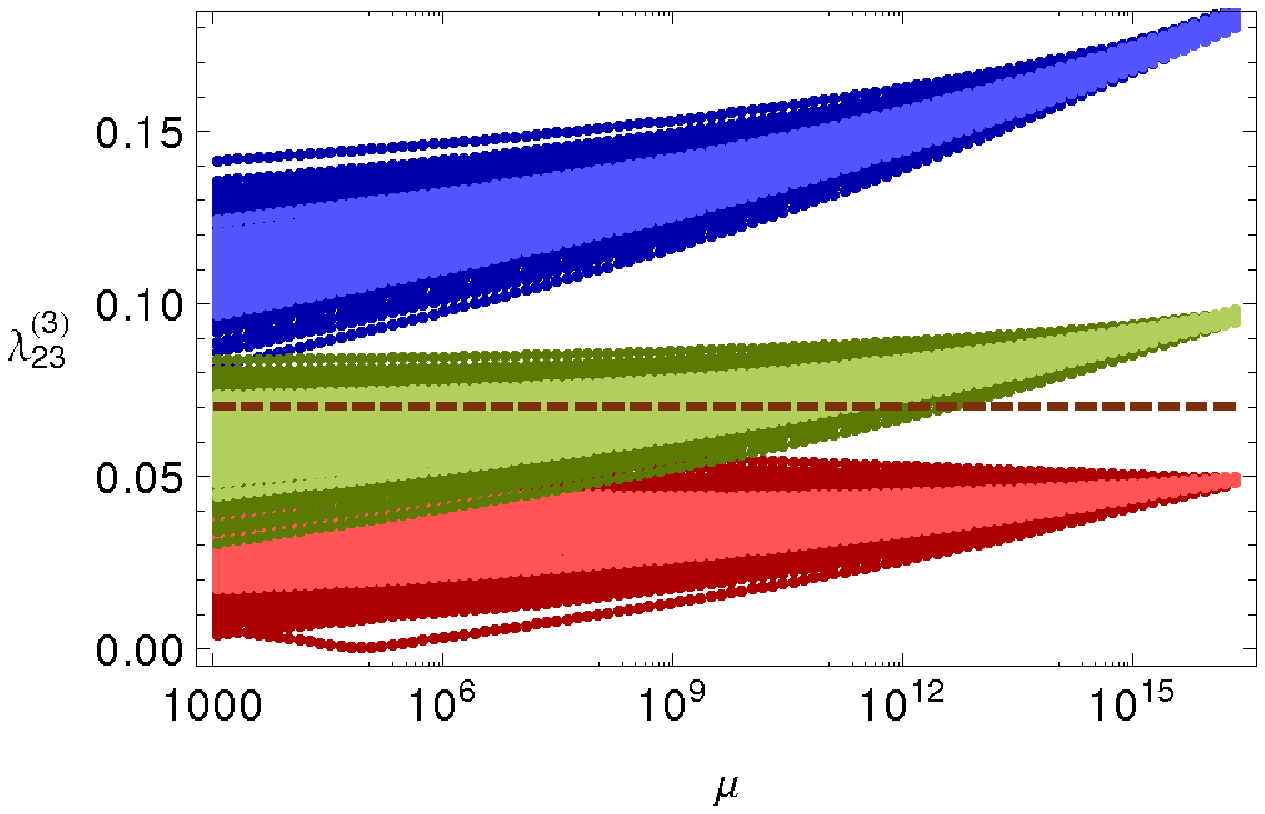} \qquad
\includegraphics[width=0.45\textwidth]{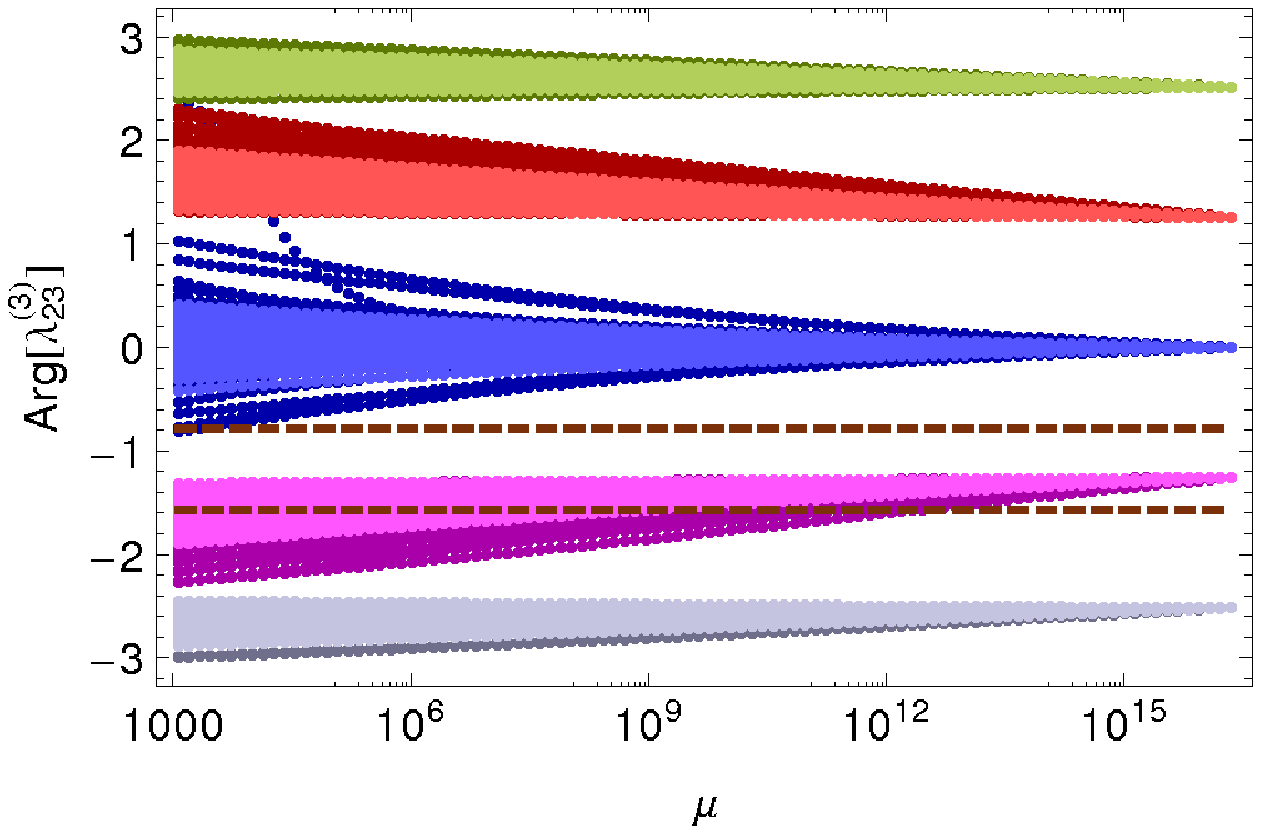} 
\end{center}
\caption{The running of $|\lambda_{23}^{(3)}|$ (left) and Arg$(\lambda_{23}^{(3)})$ (right), in Benchmark 1. On the left, we show $x_L=2,\,1,\,0.5$ in blue, green and red. In every region the lighter colour correspond to $\gamma_L$ fixed to $\pi/4$. On the right we fix $\gamma_L=(-1+0.4n)\pi$, with $n=0,\,1,\,2,\,3,\,4$ in blue, red, green, gray and magenta, respectively. The lighter regions correspond to $x_L=1$. In the first two plots the dashed brown lines mark the region where the flavour tension can be solved. On the left, the region is above the line, while and on the right it is between the two lines.}
\label{fig:lambdasB1}
\end{figure}

\begin{figure}[tbp]
\begin{center}
\includegraphics[width=0.45\textwidth]{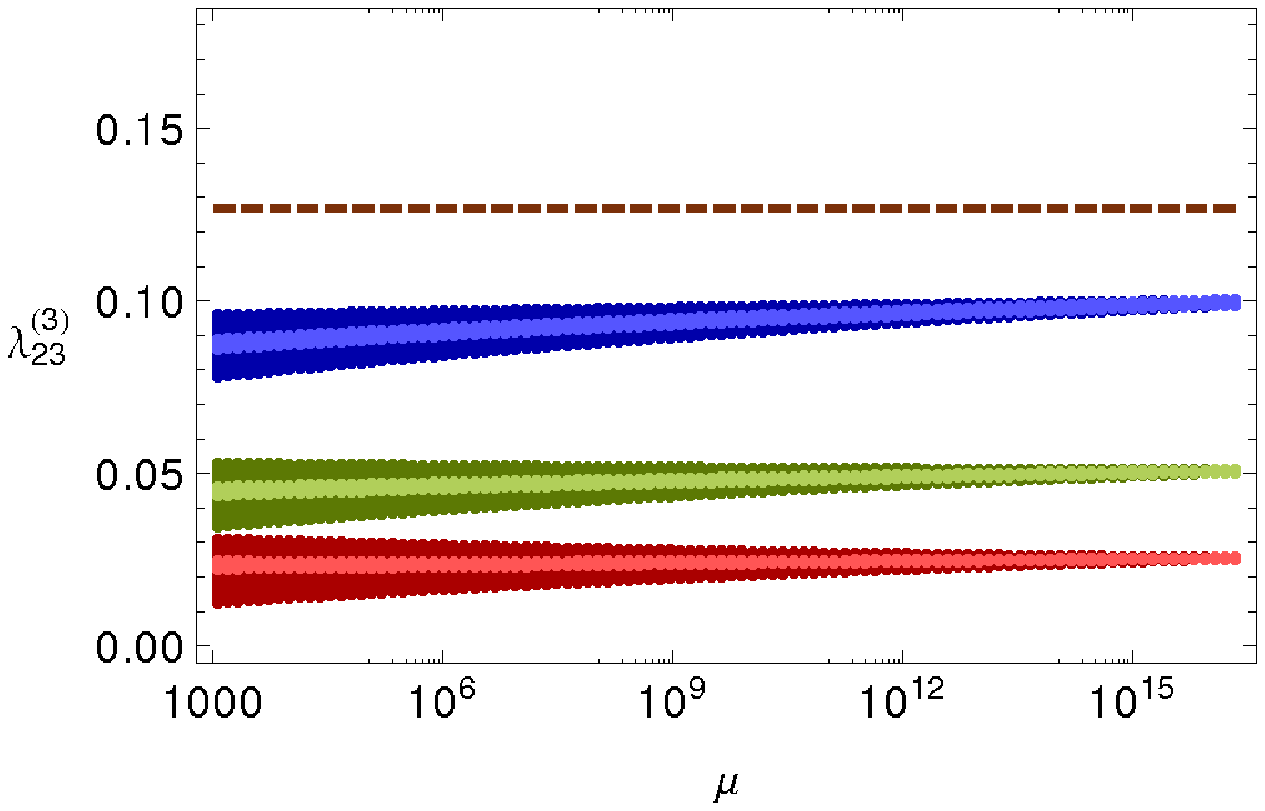} \qquad
\includegraphics[width=0.45\textwidth]{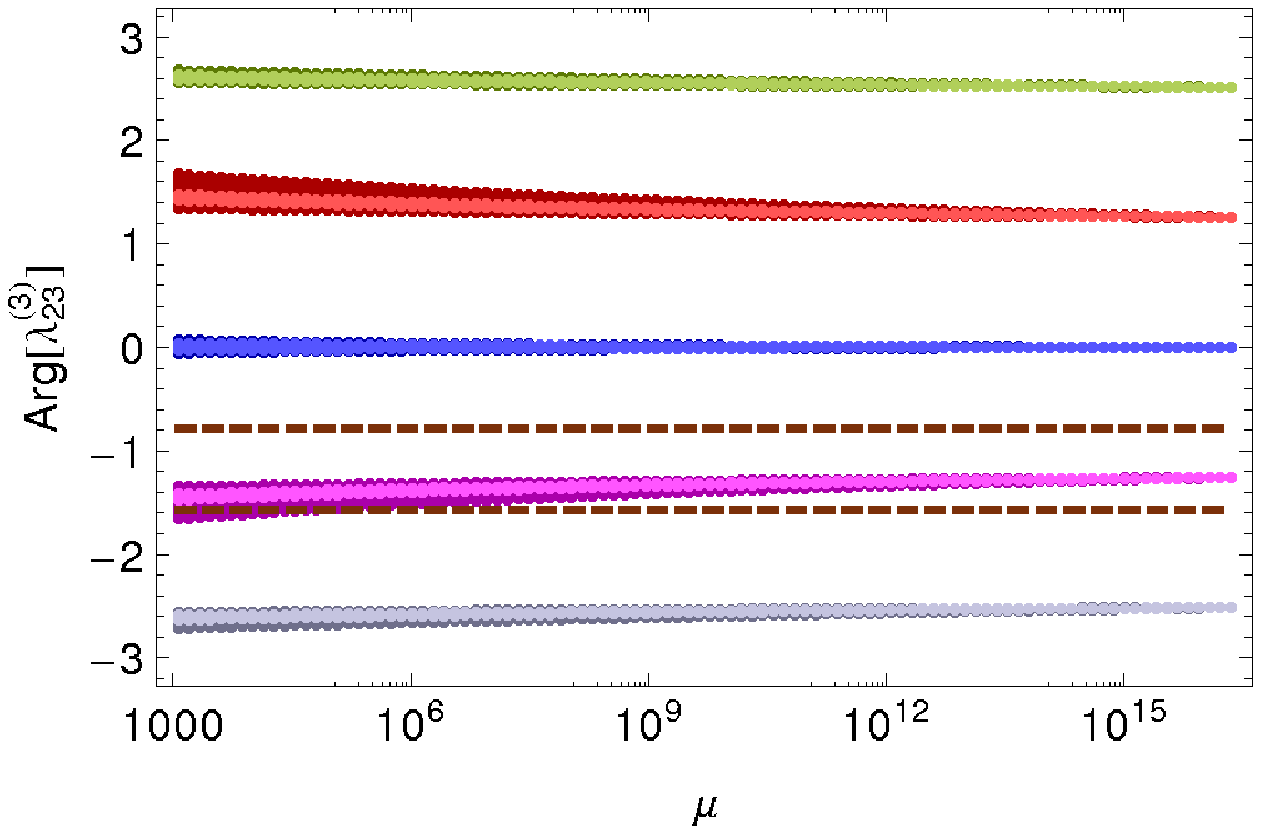} 
\end{center}
\caption{The running of $|\lambda_{23}^{(3)}|$ (left) and Arg$(\lambda_{23}^{(3)})$ (right), in Benchmark 2. Notation as in Fig.~\ref{fig:lambdasB1}. } 
\label{fig:lambdasB2}
\end{figure}

Our procedure consists in the study of the evolution of $\lambda^{(3)}_{23}$ as a function of the renormalization scale for the two Benchmarks identified in the previous Section, similarly to~\cite{Paradisi:2008qh} for MFV. In Figure~\ref{fig:lambdasB1} and~\ref{fig:lambdasB2} we plot separately the absolute value (on the left) and the phase (on the right) of $\lambda^{(3)}_{23}$ in Benchmark 1 and 2, respectively. 
For the absolute values, we fix the parameter $x_L$ that defines the mixing at $M_{\rm GUT}$ and we make a numerical scan of the other mixing parameters of the framework. 
We show with different colours three different values of $x_L$, as indicated in Figure~\ref{fig:lambdasB1}.
We also show with a lighter colour the case in which the phase $\gamma_L$ is fixed equal to $\pi/4$.
On the contrary, in the plots of the phases, each colour represents a different initial value for $\gamma_L$, varying all the other flavour parameters. Here, the lighter colour corresponds to $x_L=1$.

The main results are that, in general, the modulus and phase of $\lambda^{(3)}_{23}$ are relatively stable during the running. This is more true in Benchmark 2 than in Benchmark 1, where the running effects are stronger and the absolute values get a slight suppression. For the phases, we see a very mild spread in Benchmark 1. Moreover, it is interesting to see that it is possible to obtain a sizable phase even when starting from a real case at $M_{\rm GUT}$. This is due mainly to the influence of phases in the trilinear parameters.

In each Figure, we also mark with a brown, dashed line the region where the mixing has got the appropriate size in order to solve the flavour tension. In the previous Section, we have outlined the region of the parameter space where the function $F_0$ is large enough to solve the tension. In particular, we showed that Benchmark 1 is within this region, while Benchmark 2 is not. Nevertheless, what really solves the flavour tension is the combination $x\,F_0$, where $x$ is defined as $x={s_L^2 c^2_d/|V_{ts}|^2}$. In principle, it is possible to have a very small value of $F_0$, but a very large value of $x$, and achieve the same results as with moderate values of both parameters. In contrast, it is possible to have an appropriate value of $F_0$ and end with a too small or too big $x$.

For the absolute value, we have $x\gtrsim3$ in Benchmark 1, while $x\gtrsim10$ in Benchmark 2. The dashed lines on the respective Figures show this lower bound. For Benchmark 1, we find that values of $x_L$ of $\ord{1}$ naturally reproduce the required mixing, as long as they are greater than unity. On the other hand, for Benchmark 2, we require the initial value of $\lambda^{(3)}_{23}$ to be somewhat large in order to obtain the minimum amount of mixing. Still, it is encouraging to note that the needed initial value is not many orders of magnitude larger, such that it could be obtained at $M_{\rm GUT}$ through an accidental enhancement.

The phase of $\lambda^{(3)}_{23}$ also needs to acquire particular values. The correct values are delimited by brown, dashed lines in the respective Figures. In both scenarios we see that, since the phase variation is not too stong, it suffices to choose $\gamma_L(M_{\rm GUT})\sim\gamma_L(M_{\rm SUSY})$.

\subsection{Structure}
\label{sec:u2.structure}

So far, we have found regions within our parameter space satisfying all our requirements, with the exception of Benchmark 2 satisfying the $F_0$ constraint. We have also demonstrated that the $\lambda^{(3)}_{23}$ parameter is stable during the running and, for Benchmark 1, we have found that typical values are effectively within the ballpark that can solve the flavour tension.

Nevertheless, we have not demonstrated that the $U(2)^3$ properties are maintained after the running. In principle, even if the $F_0$ constraint is satisfied and we have the $\lambda^{(3)}_{23}$ parameter stable, it is not evident that the full set of parameters shall evolve in a way that the correlations between their contributions to the $K$, $B_d$ and $B_s$ sector are preserved. In particular, 
we know that the relations in Eq.~(\ref{eq:lambda}) that hold at $M_{\rm GUT}$ are one of the main features of this framework.
In order to check whether these relations are followed throughout the running, we shall use the following ratios:
\begin{align}
 \frac{\lambda^{(3)}_{13}}{\lambda^{(3)}_{23}} = \frac{\kappa^*}{c_d}~, & & 
 \frac{\lambda^{(3)}_{12}}{|\lambda^{(3)}_{13}|^2} = \frac{c_d}{\kappa}~,
\end{align}
which should remain valid for any value of the scale. The first ratio tests the correlations between the $B_d$ and $B_s$ sectors, while the second ratio tests those between the $K$ and $B_d$ sectors. Thus, if we find these ratios to hold within their theoretical errors, we shall consider the $U(2)^3$ symmetry to be preserved by the running.

We need to derive an approximate theoretical error for each ratio. For the absolute value of both ratios, we have found they are held within NLO corrections dependent on the value of $\rho$, which can lead to an error of at most $4\%$. For the phase, we find a fixed correction of the order of $\varphi_c=\arg(c_u c_d+s_u s_d e^{-i\phi})\approx0.02$. These considerations lead us to the following requirements in order to keep the $U(2)^3$ symmetry:
\begin{align}
 \left|\frac{\lambda^{(3)}_{13}}{\lambda^{(3)}_{23}}\right| = \left|\frac{V_{td}}{V_{ts}}\right|(1\pm0.04)~, & & 
\arg\left(\frac{\lambda^{(3)}_{13}}{\lambda^{(3)}_{23}}\right) = -\arg\left(\frac{V_{td}}{V_{ts}}e^{i\varphi_c}\right)\pm0.02~, \\
 \left|\frac{\lambda^{(3)}_{12}}{|\lambda^{(3)}_{13}|^2}\right| = \left|\frac{V_{ts}}{V_{td}}\right|(1\pm0.04)~,  & & 
\arg\left(\frac{\lambda^{(3)}_{12}}{|\lambda^{(3)}_{13}|^2}\right) = -\arg\left(\frac{V_{td}}{V_{ts}}e^{i\varphi_c}\right)\pm0.02~. 
\end{align}

\begin{figure}[tb]
\begin{center}
\includegraphics[width=0.45\textwidth]{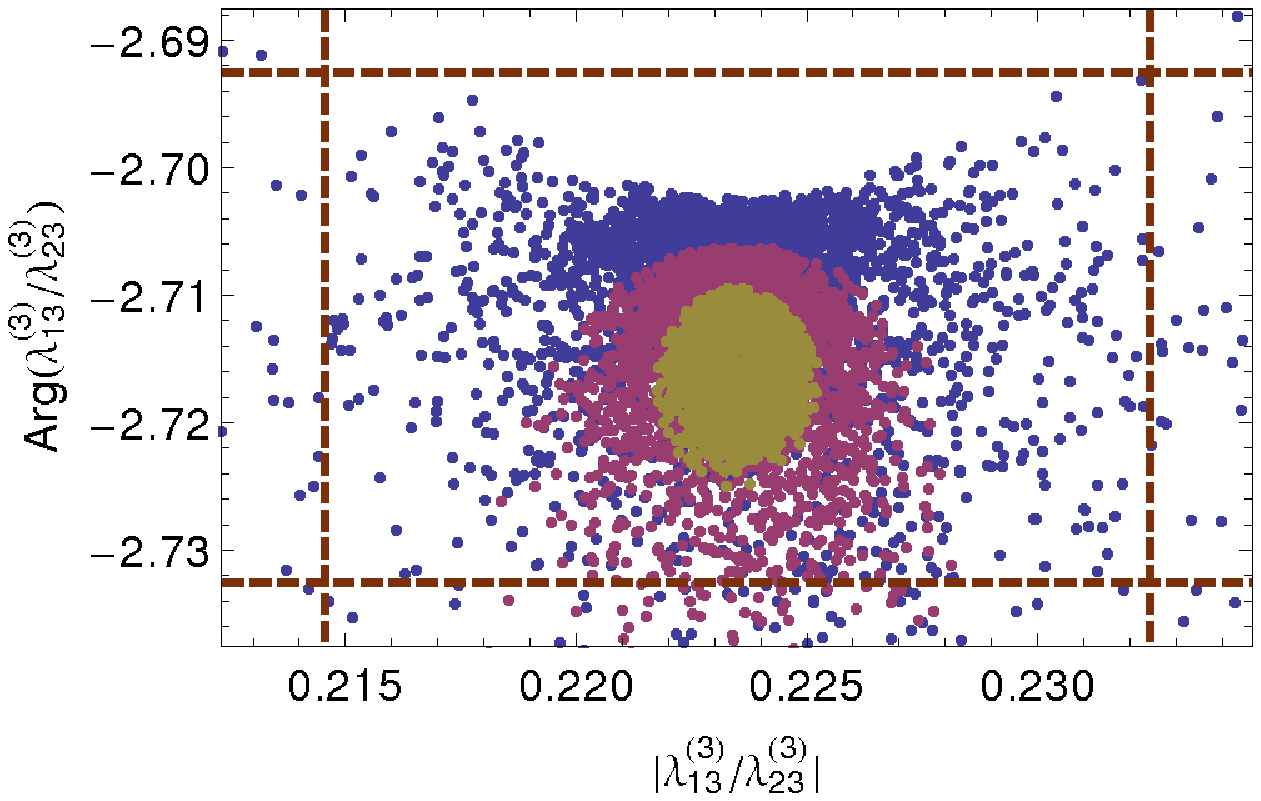} \qquad
\includegraphics[width=0.45\textwidth]{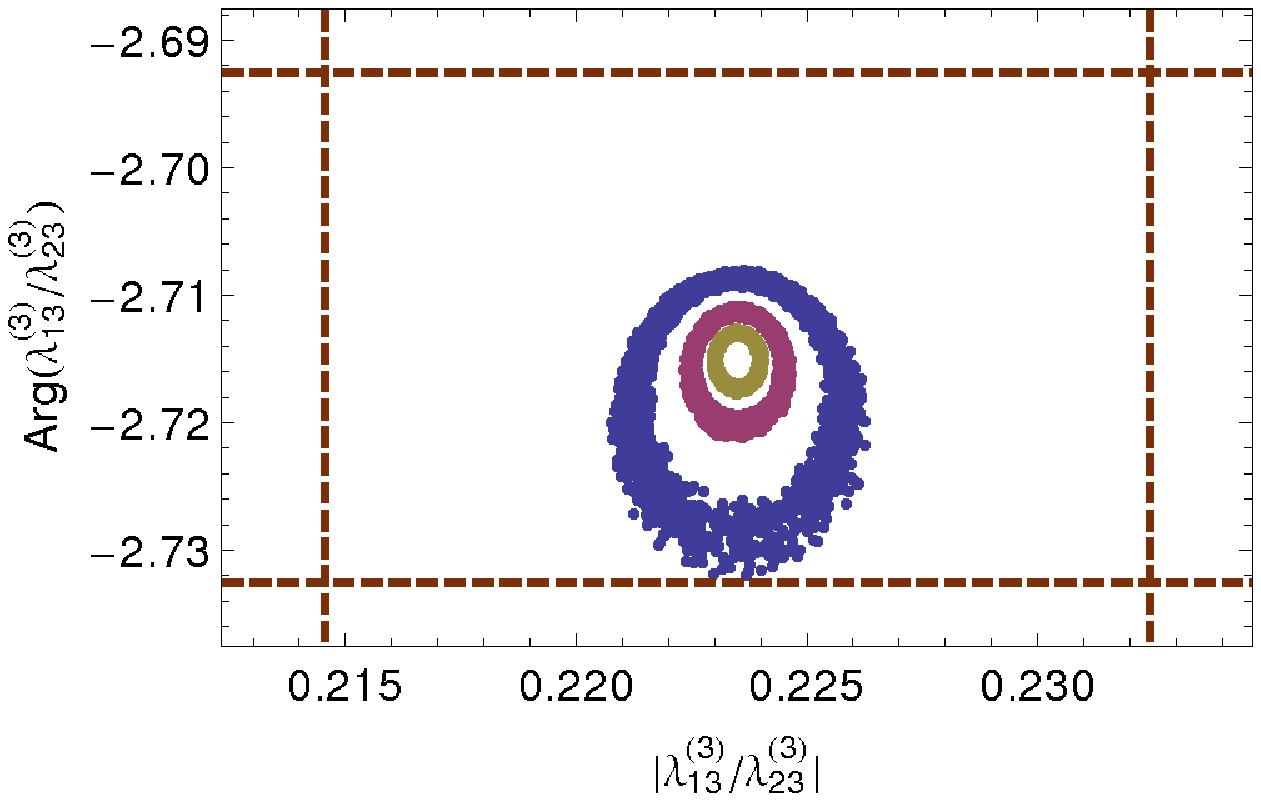} \\
\includegraphics[width=0.45\textwidth]{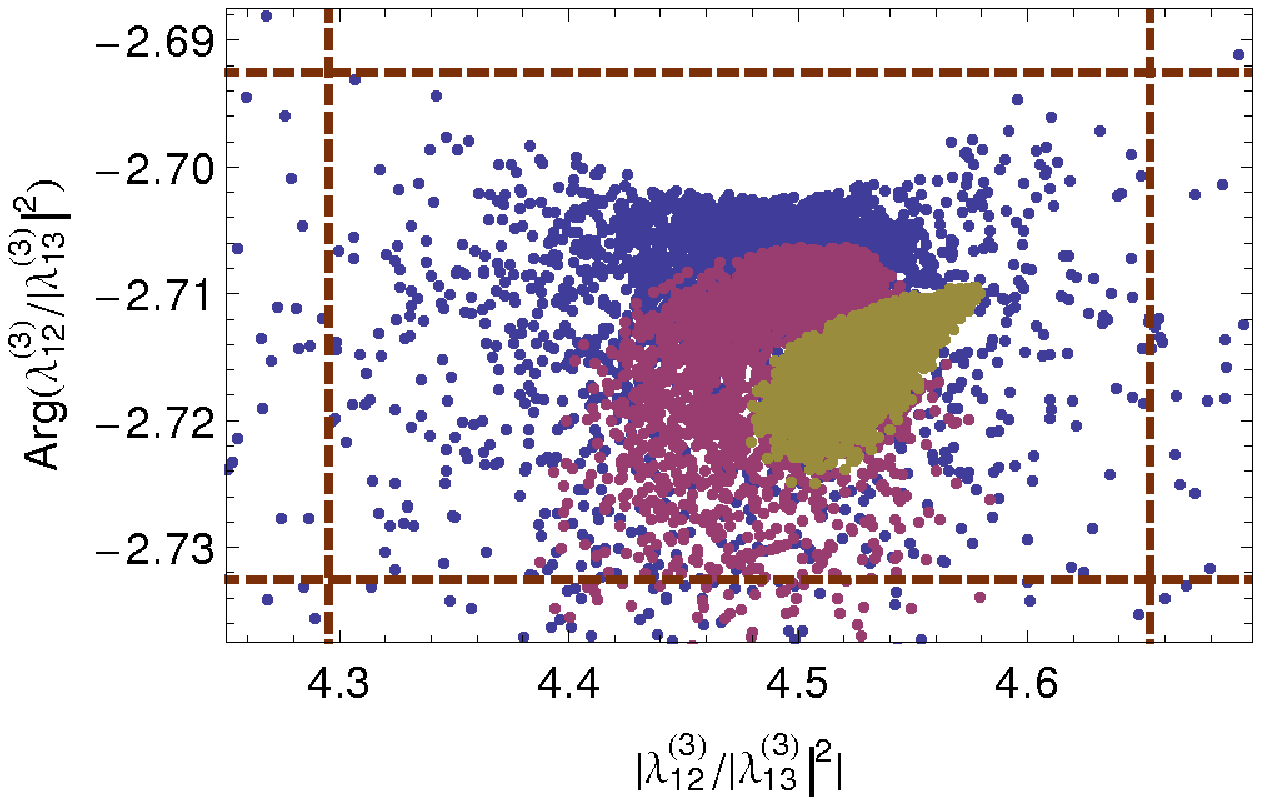} \qquad
\includegraphics[width=0.45\textwidth]{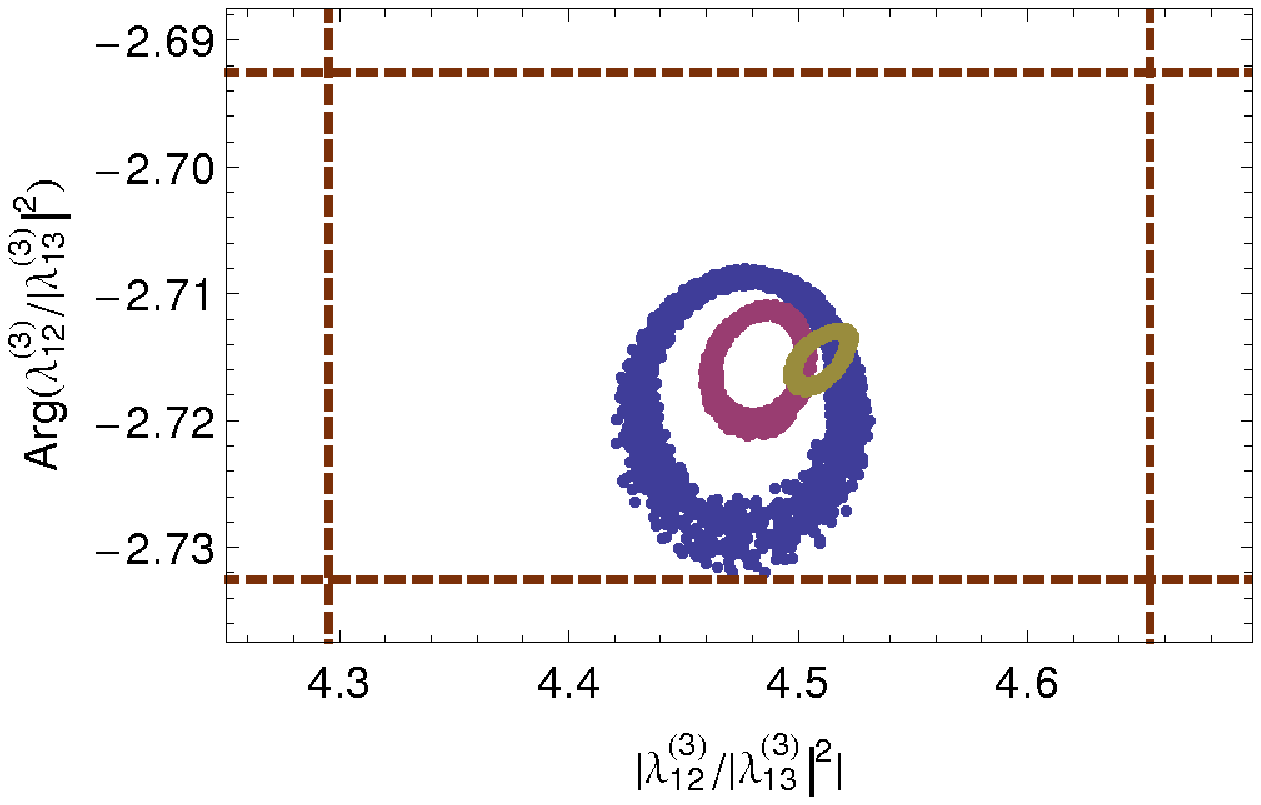}
\end{center}
\caption{The two ratios used in order to test $U(2)^3$, evaluated at $M_{\rm SUSY}$. We show Benchmark 1 (Benchmark 2) on the left (right). The dashed lines correspond to our estimated theoretical uncertainty. We show results for $x_L=2,\,1,\,0.5$ in brown, magenta and blue, respectively.} 
\label{fig:invariant}
\end{figure}
We present our results in Figure~\ref{fig:invariant}. The left (right) column shows our results for Benchmark 1 (Benchmark 2), and the top (bottom) row shows the modulus and the phase of the $\lambda^{(3)}_{13}/\lambda^{(3)}_{23}$ ($\lambda^{(3)}_{12}/|\lambda^{(3)}_{13}|^2$) ratio. We show in brown, magenta and blue the value of each ratio at $M_{\rm SUSY}$, fixing $x_L=2,\,1$ and $0.5$, respectively. The main conclusion from all plots is that the RGE variation keeps the ratios within our estimated theoretical uncertainties, so we can expect the $U(2)^3$ symmetry correlations to be preserved at all scales. Furthermore, we expect the correlations to be better mantained the larger the value of $x_L$, which is compatible with the requirement of a large $x_L$ needed to solve the flavour tension.

The distribution of points in Figure~\ref{fig:invariant} deserves an explanation, in particular for Benchmark 2, which shows a ring-like pattern. In this case, we find the pattern to be due to fixed RGE contributions, coming from the irreducible MFV terms and off-diagonal soft terms, which are of the same order. Here, the only significant variable is the effective phase between the two contributions, which is identical in all sectors, and shapes the rings.

In contrast, in Benchmark 1, we have an additional contribution from the RGEs coming from the A-Terms, which are larger than in Benchmark 2. This additional contribution involves new varying $\ord{1}$ parameters and new phases, which spoil the ring-like pattern.

\subsection{Evaluation of Operators leading to $\Delta F=2$ Processes}
\label{sec:u2.othoperator}

As mentioned previously, in $U(2)^3$ supersymmetric frameworks, the main deviations from MFV happen within $m^2_{\tilde Q}$. This suggests that the main contribution to $\Delta F=2$ processes should come from $(LL)^2$ operators, as other contributions would be strongly suppressed, usually by the masses of the first or second generation quarks.

In this section, we compare the value of the different operators contributing to $\Delta F=2$ processes after the RGE evolution, in order to make sure this is the case. We shall use the following basis for the effective operators:
\begin{equation}
 H_{\rm eff}^{F}=\sum_{i=1..5} C^F_i\,Q^F_i+\sum_{i=1..3} \tilde C^F_i\,\tilde Q^F_i~,
\end{equation}
where $F=K,\,B_d,\,B_s$ and:
\begin{align}
 Q^F_1 &= (\bar q^\alpha_L\gamma_\mu q'^\alpha_L)(\bar q^\beta_L\gamma^\mu q'^\beta_L)~, & \\
 Q^F_2 &= (\bar q^\alpha_R\,q'^\alpha_L)(\bar q^\beta_R\,q'^\beta_L)~, & 
 Q^F_3 = (\bar q^\alpha_R\,q'^\beta_L)(\bar q^\beta_R\,q'^\alpha_L)~, \\
 Q^F_4 &= (\bar q^\alpha_R\,q'^\alpha_L)(\bar q^\beta_L\,q'^\beta_R)~, & 
 Q^F_5 = (\bar q^\alpha_R\,q'^\beta_L)(\bar q^\beta_L\,q'^\alpha_R)~.
\end{align}
Here, the quarks $q,q'$ depend on the meson $F$. The $\tilde Q^F_i$ coefficients are equal to those without a tilde, with the exchange $L\leftrightarrow R$.

The Wilson coefficients $C_i^F,\,\tilde C_i^F$ have been calculated in many works, either exactly~\cite{Hagelin:1992tc,Donoghue:1983mx} or in the MIA~\cite{Gabbiani:1996hi,Ciuchini:1998ix}. In the following, we shall calculate the coefficients from the exact expressions, but shall use the MIA to discuss our results. The $(LL)^2$ contribution corresponds to the $C^F_1$ coefficient. Similarly, the $C^F_2$ and $C^F_3$ coefficients correspond to the $(RL)^2$ contributions, while the $C^F_4$ and $C^F_5$ coefficients correspond to $(LL)(RR)+(LR)(RL)$ contributions. Again, these are related to $\tilde C^F_i$ by the exchange $L\leftrightarrow R$.

Given the vanishing value of the lower off-diagonal elements of $A_d$ at the GUT scale, and the very small MFV contribution from the running, the $(RL)^2$ contributions are expected to be the smallest. Next in the line come the $(LR)^2$ contributions, which, although involving non-negligible upper off-diagonal elements of $A_d$, also include an additional suppression proportional to $m^2_b/m^2_{heavy}$. This shall compete with the $m^2_q/m^2_b$ suppression commonly found in the $(RR)^2$ contribution, where $m_q$ can be either the first or second generation quark mass, depending on sector involved. Finally, the $(LL)(RR)$ contribution should be the largest after the $(LL)^2$, given the relatively small suppression of the $(LL)$ insertion. Thus, from the mixing point of view, we would expect:
\begin{equation}
 C^F_2\sim C^F_3 \ll \tilde C^F_1\sim \tilde C^F_2 \sim \tilde C^F_3 \ll C^F_4\sim C^F_5 \ll C^F_1~.
\end{equation}
\begin{figure}[tbp]
\begin{center}
$K$ \\
\includegraphics[width=0.45\textwidth]{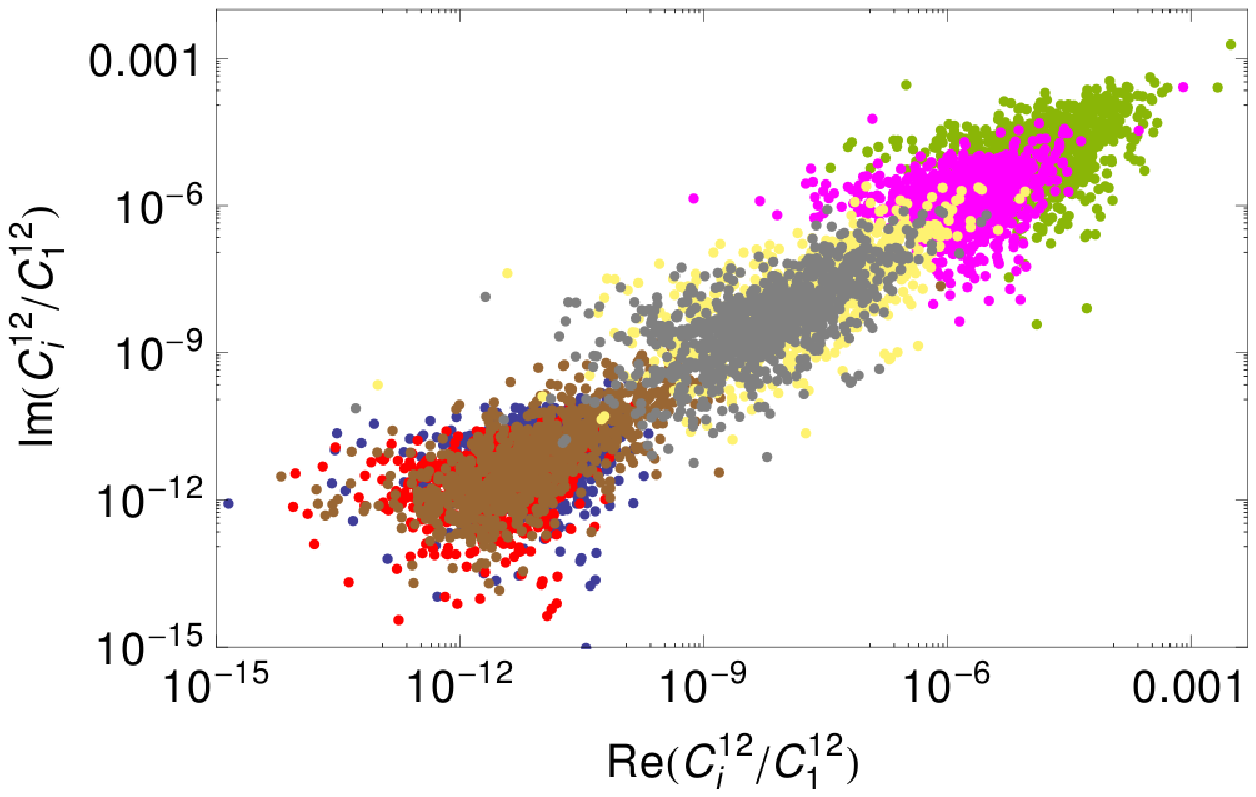} \quad
\includegraphics[width=0.45\textwidth]{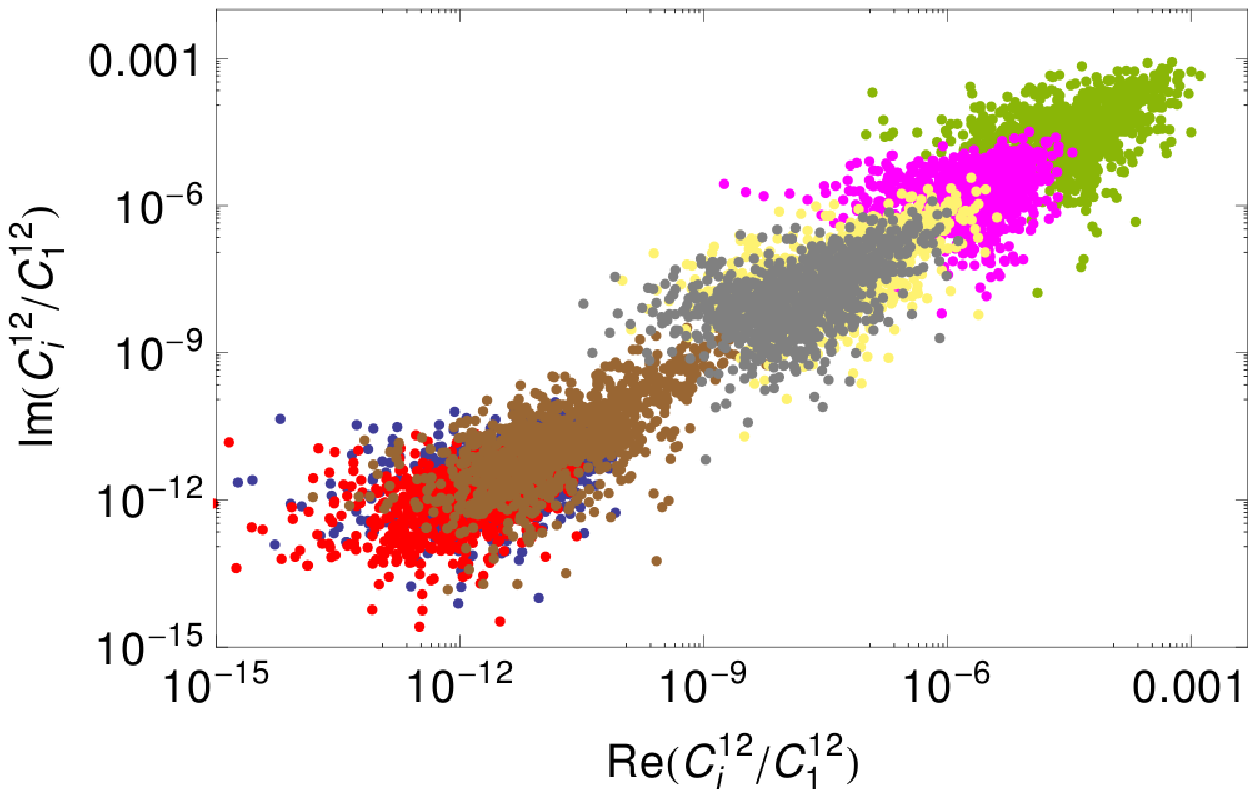} \\
$B_d$ \\
\includegraphics[width=0.45\textwidth]{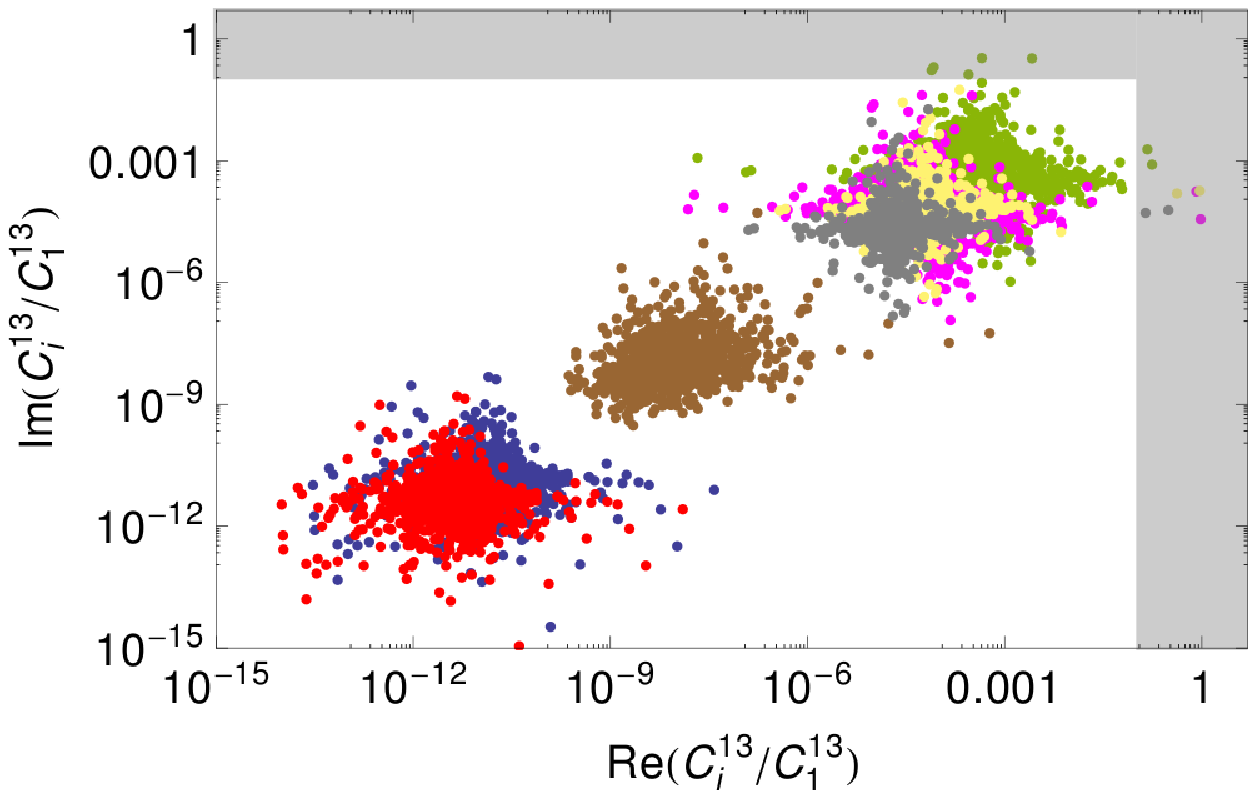} \quad
\includegraphics[width=0.45\textwidth]{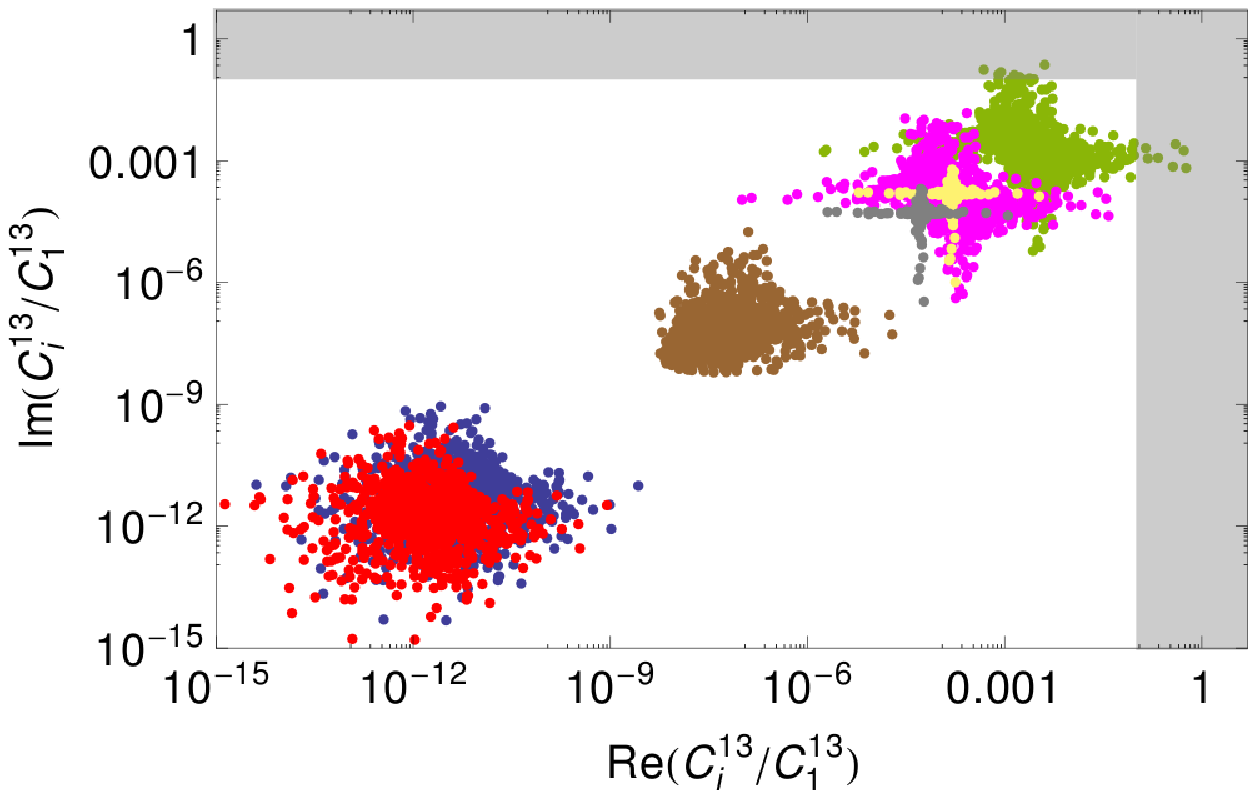} \\
$B_s$ \\
\includegraphics[width=0.45\textwidth]{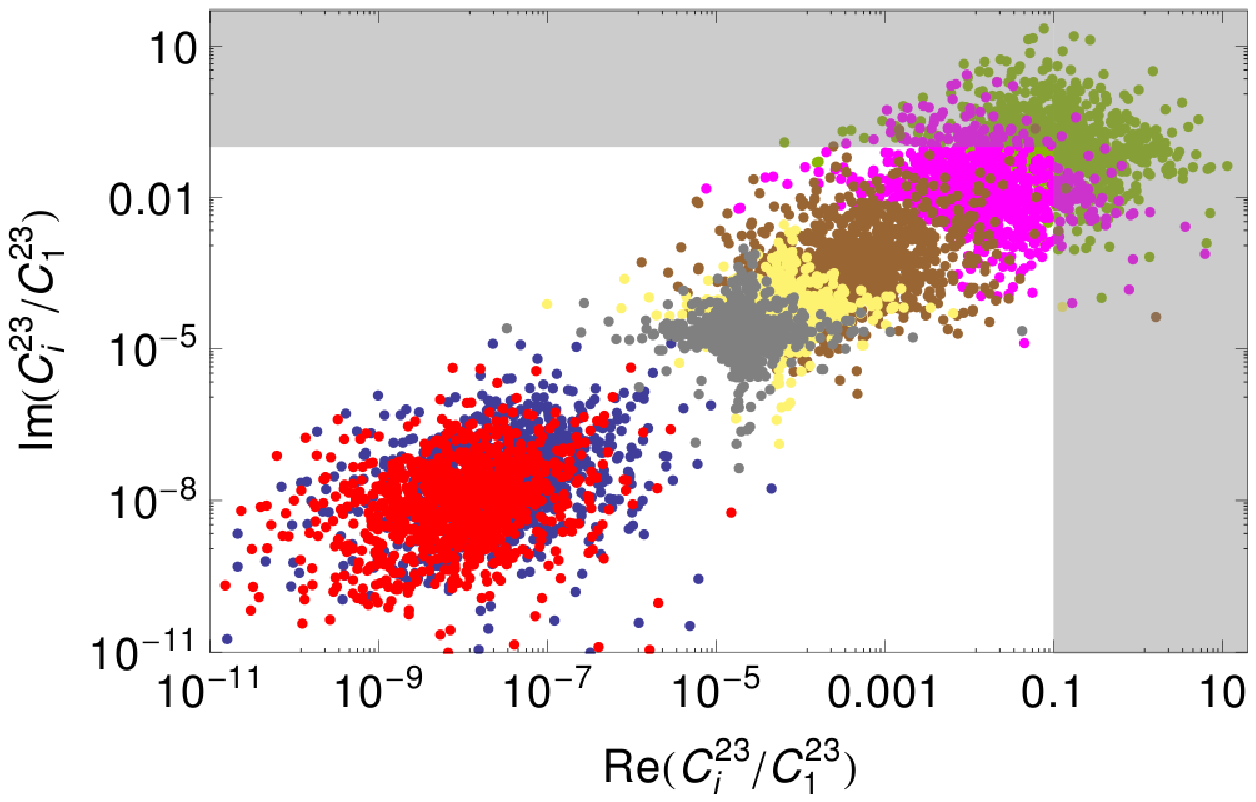} \quad
\includegraphics[width=0.45\textwidth]{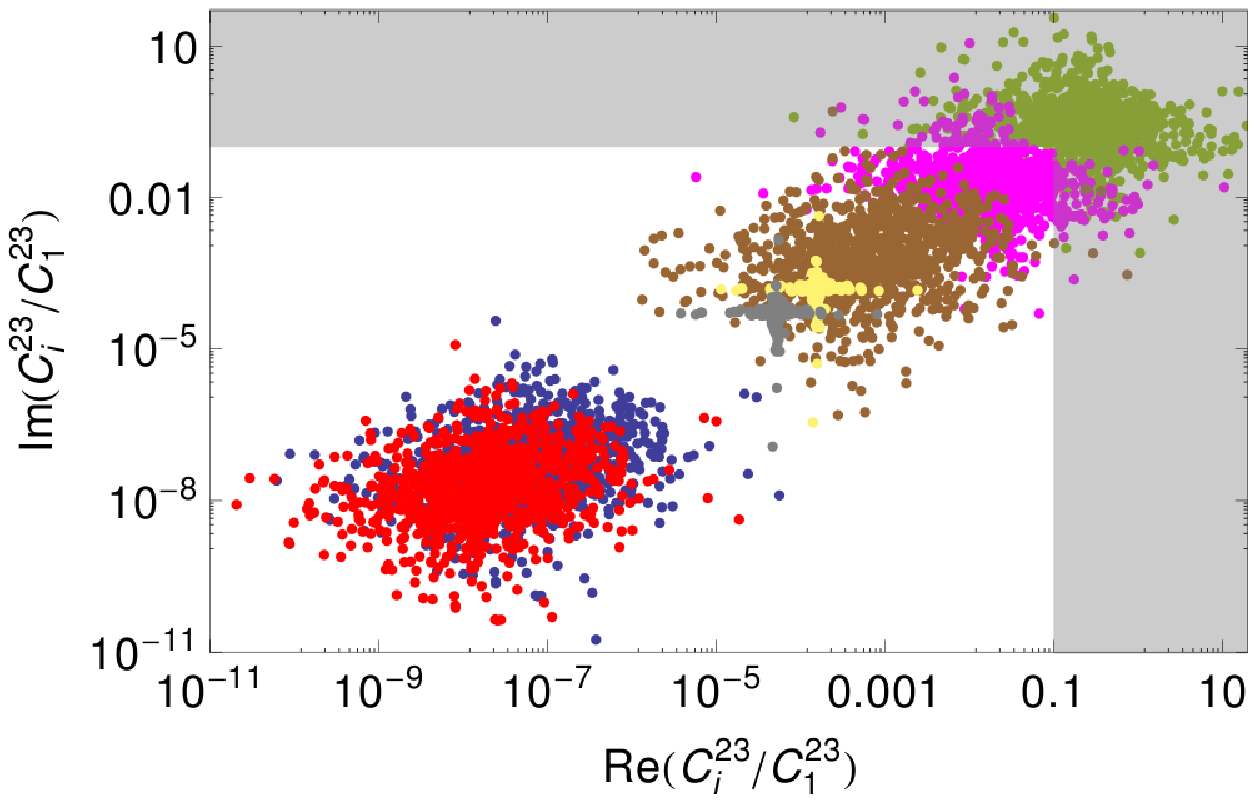}
\caption{Ratios between $C^F_i,\,\tilde C^F_i$ and $C^F_1$ coefficients. We show the ratios for the $K$, $B_d$ and $B_s$ sectors in the top, centre and bottom panels, respectively. Benchmark 1 (2) is shown on the left (right) column. The $C^F_i/C^F_1$ ratios are shown in blue, red, green and magenta, for $i=2\ldots5$. The $\tilde C^F_i/C^F_1$ ratios are shown in brown, yellow and gray for $i=1\ldots3$. The shadowed regions mark the areas where the ratios are larger than 10\%.} 
\label{fig:coefficients}
\end{center}
\end{figure}

The results are very similar to our expectations, and are shown in Figure~\ref{fig:coefficients}. Here, we show the $C^F_i/C^F_1$ and $\tilde C^F_i/C^F_1$ ratios, for all possible coefficients, in the $K$, $B_d$ and $B_s$ sectors. The coefficients are calculated at $M_{\rm SUSY}$, and for transparency are not evolved to the respective meson scale. We find the hierarchies between Benchmark 1 and 2 are identical, with a smaller spread for $\tilde C^F_2$ and $\tilde C^F_3$ in Benchmark 2, due to the additional suppression in $A_0=0.25\,m_{heavy}$.

In the Figure, the shadowed regions indicate values where the ratios exceed $10\%$, meaning they should not be neglected. Surprisingly enough, we find that in the $B_s$ sector the $C_4$ coefficient can be well within this region, and can actually become as much as ten times larger than $C_1$, especially in Benchmark 2. This might spoil the correlation between CP violation in the $B_d$ and $B_s$ sectors and, more importantly, break the invariance of $\Delta M_d/\Delta M_s$ with respect to the Standard Model values.

\begin{figure}[tbp]
\begin{center}
\includegraphics[width=0.5\textwidth]{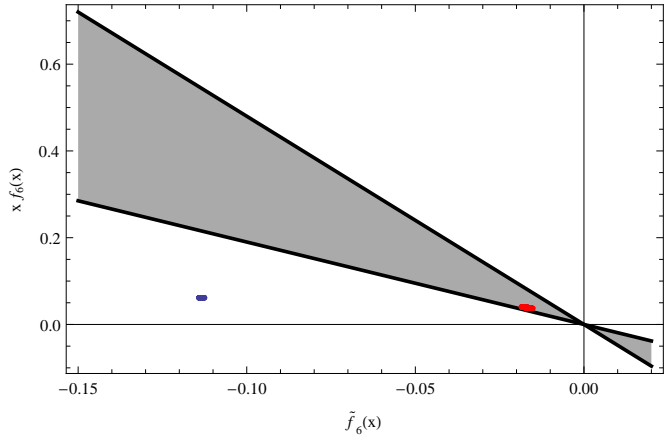}
\caption{Region where loop functions provide an additional suppression of $\ord{m_s/m_b}$ on the $(LL)^2$ operator. Typical values of Benchmark 1 (2) are shown with the blue (red) dots.} 
\label{fig:loopf}
\end{center}
\end{figure}

We have found this unexpected behaviour to be due to the small value of the loop functions for both Benchmarks, that can balance the suppression in the $RR$ mixing. This can be better understood by demanding the loop function in $C_1$ to include an additional suppression of the $\ord{m_s/m_b}$ with respect to the loop function for the dominant $(LL)(RR)$ contribution in $C_4$. This gives us:
\begin{align}
 \left|24x\,f_6(x)+66\tilde f_6(x)\right|<(m_s/m_b)\left|504x\,f_6(x)-72\tilde f_6(x)\right|
\end{align}
where $x=m^2_{\tilde g}/m^2_{\tilde b_L}$, and the loop functions $f_6(x)$ and $\tilde f_6(x)$ can be found, for instance, in~\cite{Ciuchini:1998ix}\footnote{In the mass-insertion approximation, one should actually use $x=m^2_{\tilde g}/\langle m^2_{\tilde d}\rangle$, where $\langle m^2_{\tilde d}\rangle$ is the average down squark mass~\cite{Raz:2002zx}. Nevertheless, in a split scenario this does not always give an accurate result.}. The region giving such a suppression is shown in Figure~\ref{fig:loopf}, where we can see that Benchmark 2 lies within it, while Benchmark 1 lies very close.

One finds that this small value in the loop function is actually due to the stringent bounds of the LHC on the gluino mass. In fact, in order to avoid the suppression, and have $C_4<0.1\,C_1$, one needs:
\begin{equation}
m_{\tilde{b}_L} > 3.2\,\, m_{\tilde{g}} \,,
\end{equation}
apart from $O(1)$ coefficients. Considering the LHC limit of $m_{\tilde{g}}\gtrsim1$ TeV, this bound is incompatible with a split scenario in which the third generation is relatively light. In Benchmark 2, the gluinos are heavier than in Benchmark 1, while the the sbottom are always around 2 TeV. Thus, the enhancement in $C_4/C_1$ is usually stronger.

However, we have checked that the regions where the $(LL)(RR)$ operators dominate are those where the $\ord{1}$s of the $(LL)^2$ are small. In these cases, the contributions to flavour shall always be negligible, meaning that, if this framework can solve the flavour problem, the $(LL)(RR)$ operators shall contribute at about $10\%$ of the total SUSY contribution. 

\section{$U(2)^3$ as a broken subgroup of $U(3)^3$}
\label{sec:u2dev}


Recently, an extension of this framework for the lepton sector has been presented~\cite{Blankenburg:2012nx}. It was found that, in order to reproduce the neutrino oscillation data, it was necessary to enlarge the symmetry to $U(3)^5$, i.e.\ to restore MFV. A two-step breaking would then be carried out. In the first step, we would have a breaking in two directions: one preserving $O(3)_L$ in the neutrino sector, and another one preserving $U(2)^5$ in the Yukawa sector. This would be followed by a sub-leading hierarchical breaking of $U(2)^5$, leading to the Yukawa matrices studied in this paper. At the same time, this sub-leading breaking would be connected to the neutrino sector, reproducing the observed neutrino oscillation parameters.

In this case, to introduce the $U(2)_L$ doublet, the embedding in $U(3)^5$ would force the use of a spurion transforming as an {\bf 8} of $U(3)_L$. In $U(2)^5$ language, this would have the effect of having, in addition to the usual $U(2)_L$ doublet, a new spurion $\Delta_L$, transforming with the adjoint of $U(2)_L$. Both spurions would be contained in the same representation of $U(3)_L$. 

Following this breaking, we study the effects of the corresponding spurion $\Delta_Q$, transforming as a {\bf 3} of $U(2)_Q$, in the quark sector. This modification does not alter the Yukawa structure, and affects only the $(1-2)$ block of $m_{\tilde{Q}}^2$. In particular, Eq.~(\ref{mq.spur}) in the Appendix is modified to
\begin{equation}
\frac{m^2_{\tilde Q}}{m^2_{Q_h}} =  I+
\left(\begin{array}{c:c} 
c_Q\, \Delta_Q + c_{Qv}\,V^* V^T   +  c_{Qu}  \Delta Y_u^* \Delta Y_u^{T} 
+  c_{Qd}  \Delta Y_d^* \Delta Y_d^{T} 
&   x_{Q}\, e^{-i\phi_Q} V^* \\ \hdashline
    x_{Q}\, e^{i\phi_Q} V^T   &   -\rho_Q 
\end{array}\right)~.
\end{equation}

Note that, even without considering the leptonic case, it is of general interest to study if other non-minimal breakings of $U(2)^3$ can be compatible with low energy data. In this case, the addition of the new spurion affects only the soft sector, and would be a further deviation from MFV with new physics effects in the $K$ sector. This is particularly relevant if the first two generations of squarks are not too heavy.

The most important constraints in $K$ sector come from $\epsilon_K$ and $\Delta\,M_K$, which can get an additional contribution from gluino-mediated processes involving only the first two generation squarks. This contribution is negligible in the minimal $U(2)^3$ breaking, since the 1-2 mixing has got a strong MFV suppression. On the contrary it can be sizable with the $\Delta_Q$ spurion, for example in Benchmark 1, where $m_{heavy}\simeq 3$ TeV. In the following, we shall refer to the SUSY contribution to $\epsilon_K$ coming exclusively from the $\Delta_Q$ spurion as $(\epsilon_K)_{12}$.\footnote{For a detailed analysis of the full contribution to $\epsilon_K$ in natural SUSY, see~\cite{Mescia:2012fg}.}

For simplicity we first assume that all the elements of $\Delta_Q$ are of the same size 
\begin{equation}
\label{eq:DeltaQ1}
\Delta_Q = \epsilon' \left(
\begin{array}{cc}
1 & e^{i \gamma_{12}} \\ e^{-i \gamma_{12}}& 1
\end{array}\right)~,
\end{equation}
neglecting the contributions of the other spurions to the $(1-2)$ block.

\begin{figure}[tbp]
\begin{center}
\includegraphics[width=0.4\textwidth]{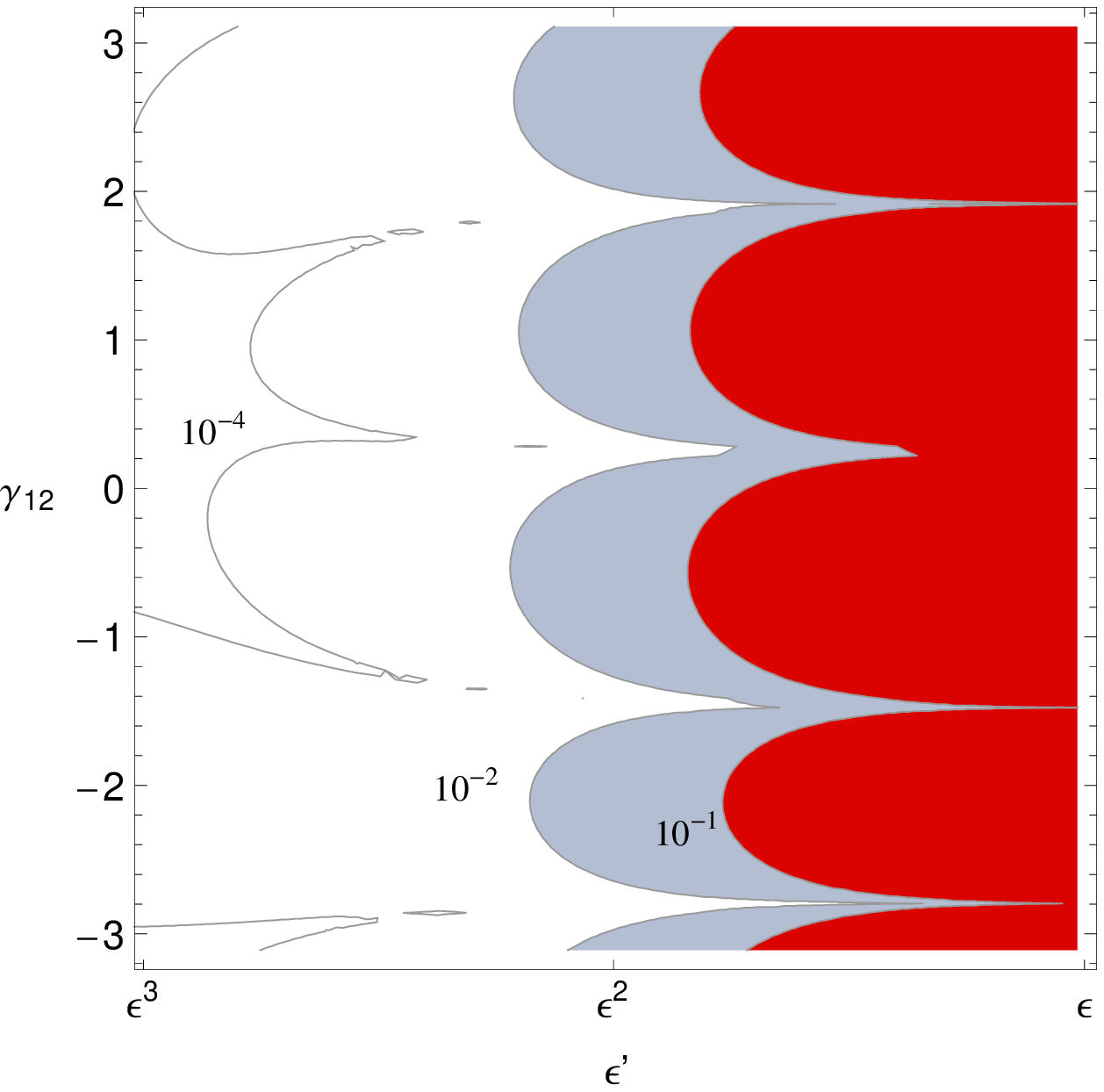} \qquad
\includegraphics[width=0.4\textwidth]{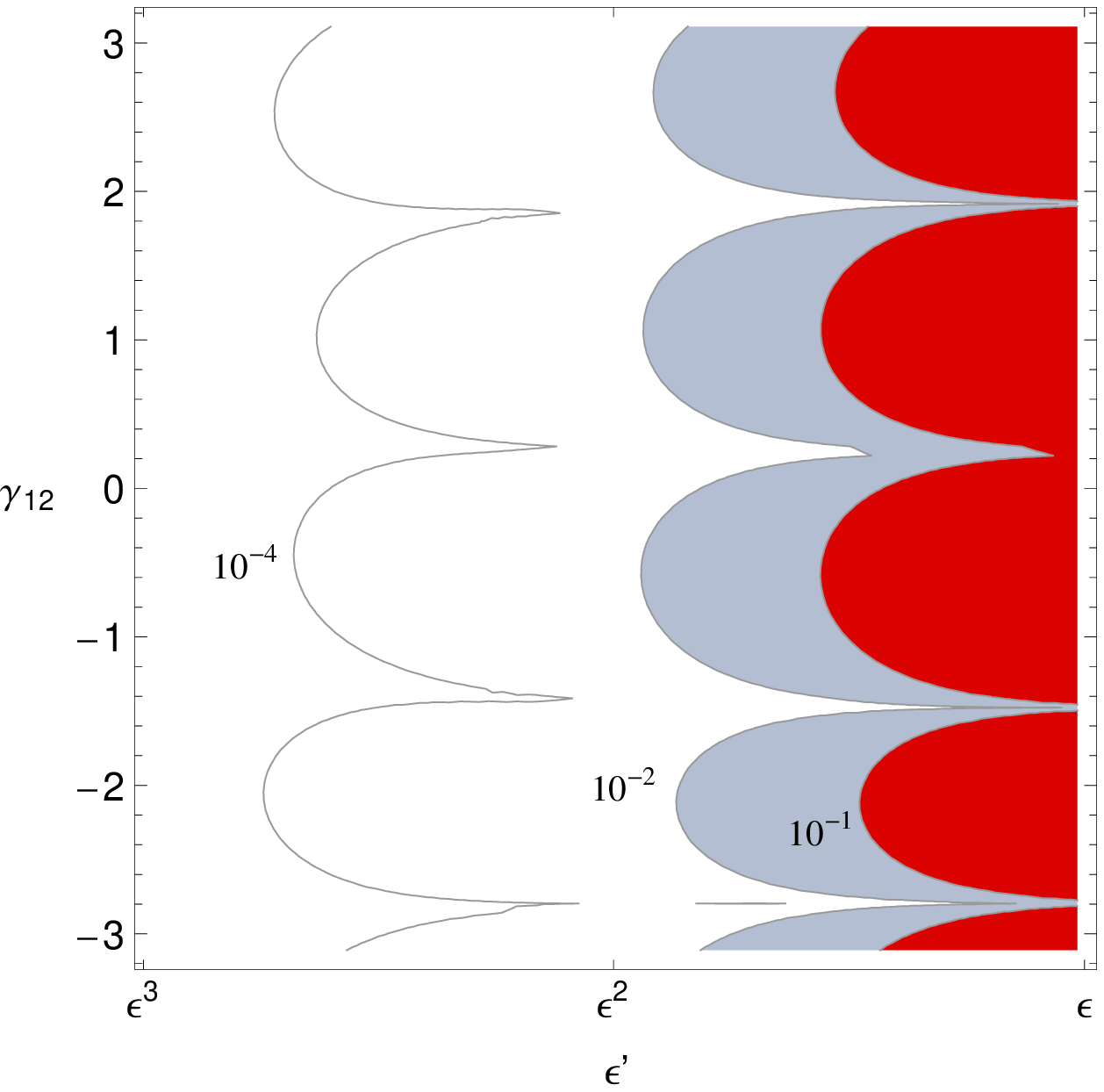}
\end{center}
\caption{The effects of the $\Delta_Q$ spurion on $K$ physics in Benchmark 1 (on the left) and 2 (on the right). We show the ratios $(\epsilon_K)_{12}/\epsilon_K^{exp}$ as a function of the $\Delta_Q$ elements, following Eq.~(\ref{eq:DeltaQ1}). We show in red (gray) the region where $(\epsilon_K)_{12}<0.1\, \epsilon_K^{exp}$ ($(\epsilon_K)_{12}<0.01\, \epsilon_K^{exp}$).}
\label{fig:DeltaQ1}
\end{figure}
In Figure~\ref{fig:DeltaQ1} we show the contours for the ratio of $(\epsilon_K)_{12}$ on the experimental values, as a function of $\epsilon'$ and $\gamma_{12}$ in Benchmark 1 (on the left) and 2 (on the right). We show in red the region where the new contribution is bigger than $10\%$ of the experimental value. This would mean that, given a similar error in the SM prediction~\cite{Brod:2011ty}, the new effect cannot be neglected. The gray regions, on the other hand, provide a contribution larger that $1\%$, and although not dangerous, could be important in the solution of the flavour tension. We don't show the contributions to $\Delta\,M_K$, that are always very small and provide no constraint.

The main result is that, if we want the new contributions to be smaller than $1\%$, we need $\epsilon'\lesssim\epsilon^2$ (barring fine-tuning of the phase). As expected in the Benchmark 2, where the first two generations are heavier, $\Delta_Q$ can take somewhat larger values.

\begin{figure}[tbp]
\begin{center}
\includegraphics[width=0.4\textwidth]{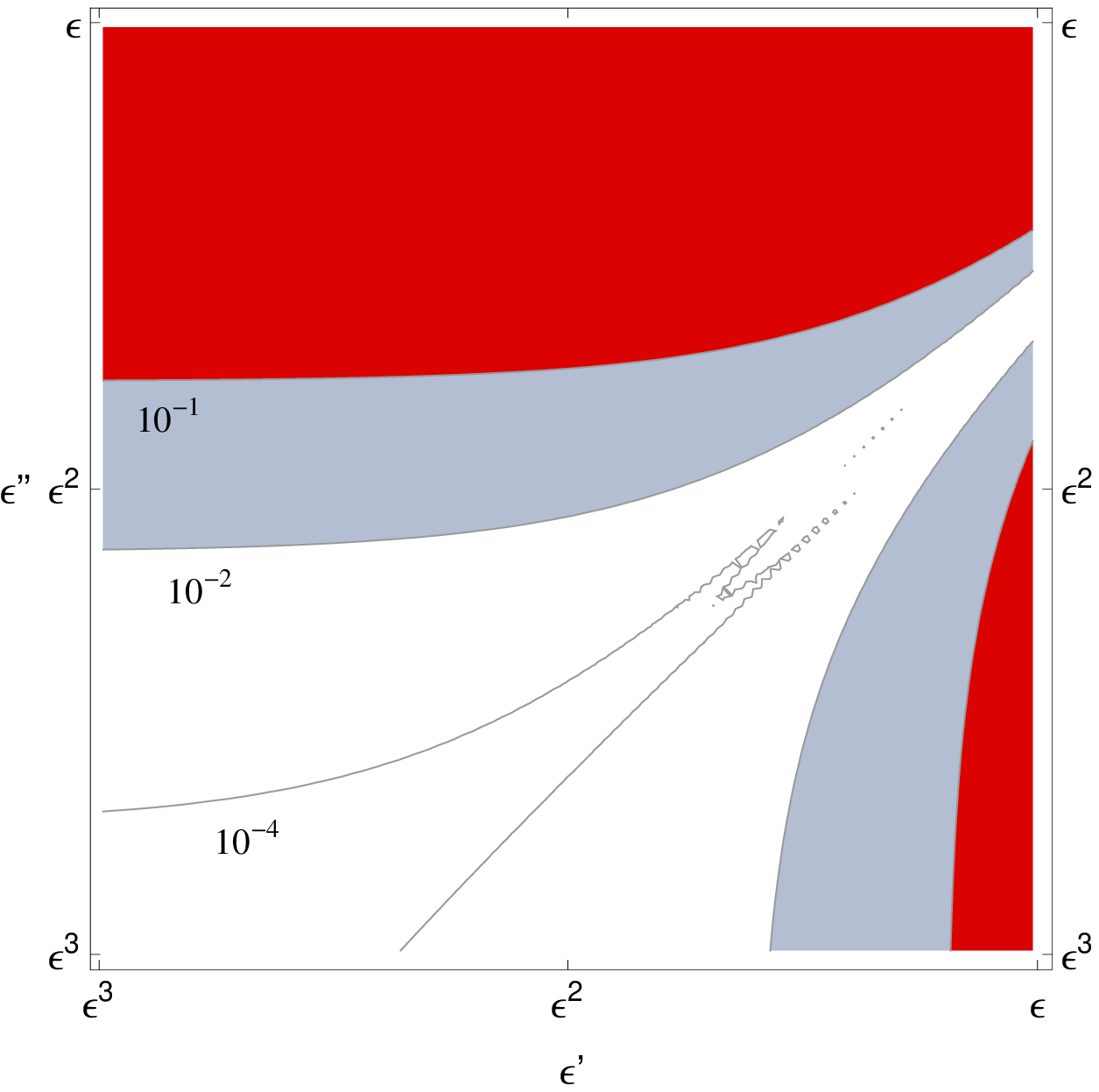} \qquad
\includegraphics[width=0.4\textwidth]{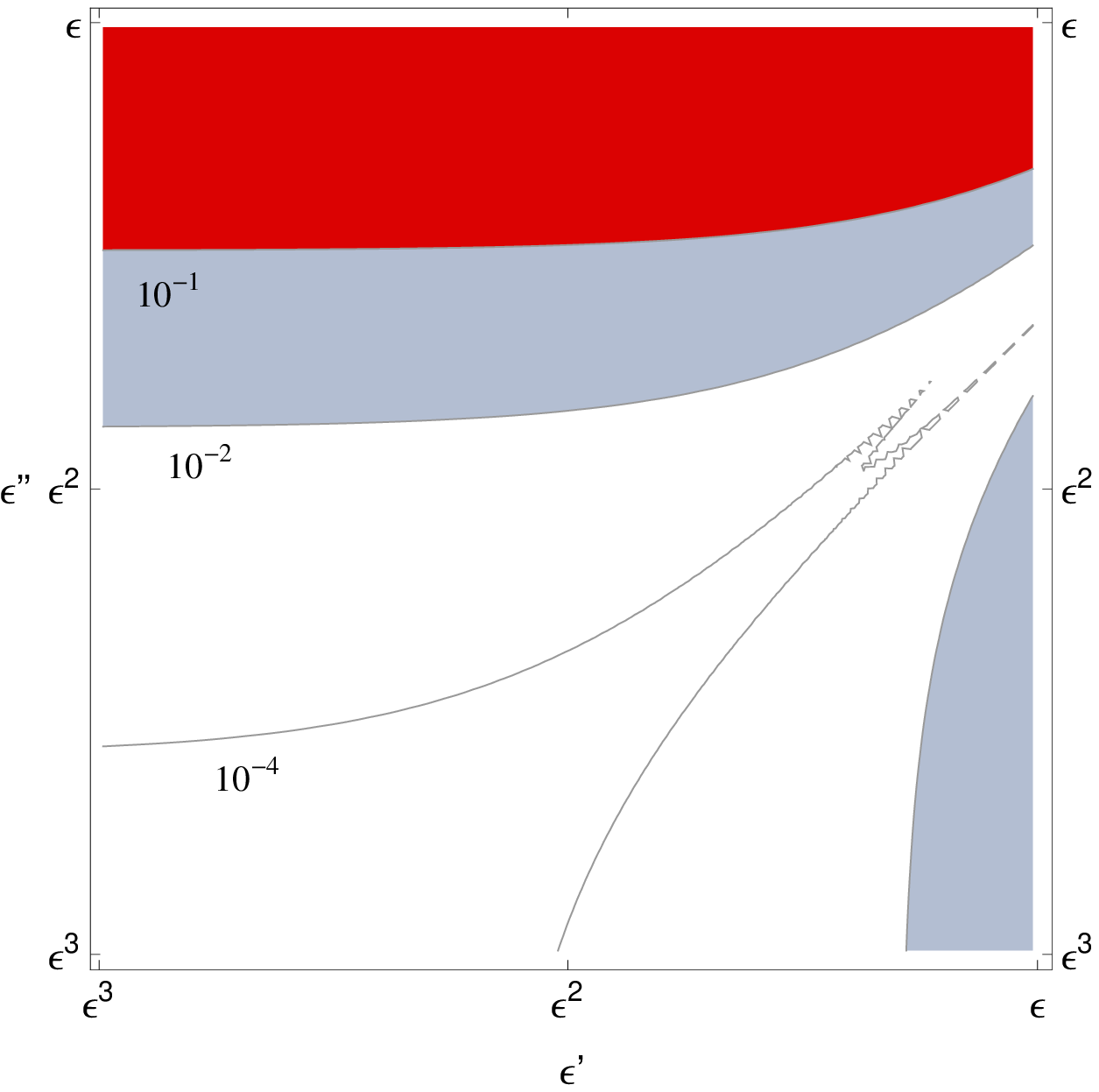}
\end{center}
\caption{The effects of the $\Delta_Q$ spurion on $K$ physics in Benchmark 1 (on the left) and 2 (on the right).We show the ratios $\epsilon_K/\epsilon_K^{exp}$ in function of the $\Delta_Q$ elements as explained in Eq.~(\ref{eq:DeltaQ2}). In red the region where $\epsilon_K<0.1\, \epsilon_K^{exp}$.}
\label{fig:DeltaQ2}
\end{figure}

Another interesting case to consider is the one with the elements of $\Delta_Q$ being of different sizes. For example, we take the case:
\begin{equation}
\label{eq:DeltaQ2}
\Delta_Q = \left(
\begin{array}{cc}
0 & \epsilon''\, e^{i \gamma_{12}} \\ \epsilon''\, e^{-i \gamma_{12}}& \epsilon'
\end{array}\right)~,
\end{equation}
which is precisely the form of the spurion introduced in \cite{Blankenburg:2012nx}. For simplicity, we fix the phase $\gamma_{12}=\pi/4$ and again show the ratio of $(\epsilon_K)_{12}$ over $\epsilon_K^{exp}$ as a function of $\epsilon'$ and $\epsilon''$ in Figure~\ref{fig:DeltaQ2}. As expected, the ratio is very small when both $\epsilon'$ and $\epsilon''$ are small, and increases accordingly with them. The impact of $\epsilon'$ on $(\epsilon_K)_{12}$ is much smaller, as it enters with an additional suppresion proportional to $s_d$. For Benchmark 1, $\epsilon''\lesssim\epsilon^2$ and $\epsilon'\lesssim\epsilon/5$ assure that the contribution to $\epsilon_K$ shall be lower than $\sim1\%$, while for Benchmark 2 the bounds again are much milder. Notice, however, there exists a region where the $\epsilon''$ contribution cancels the $\epsilon'$ contribution. This interference depends on the values of $\gamma_{12}$ and $\alpha_d$.

Moreover, we note that the RGEs effects on 1-2 sector are very weak, and the values of $\Delta_Q$ do not change significantly during the running.



\section{Conclusions}

We have studied the running behaviour of a split-family SUSY framework based on a $U(2)^3$ family symmetry. As mentioned in the introduction, such a framework is motivated by the current lack of experimental evidence for SUSY at the early runs of the LHC, and by the existence of a small flavour tension between the $K$ and $B_d$ sectors. Nevertheless, it was not evident if the several low-scale analyses of this framework were valid if the symmetry was actually broken at a large scale.

In this work, we studied the $U(2)^3$ framework through a CMSSM-like parameter space, and understood the consequences on the low-energy spectrum. This was made clear through the use of two benchmark scenarios, the first one having the heavy squarks slightly beyond the current reach of the LHC, and the second one having them considerably heavier.

Theoretical consistency, along with the requirement of reproducing the Higgs mass and solving the flavour tension at the low scale, forced Benchmark 1 to have a very specific spectrum, with a very light stop and somewhat heavier sbottoms. Here, we found that the evolution of the mixing parameters was very mild, and required the relevant $\ord{1}$ constants to be slightly larger than unity in order to successfully solve the flavour tension. The correlations between the $(LL)^2$ SUSY contributions to $K$, $B_d$ and $B_s$ physics were found to be preserved, but it was found necessary to check explicitly the magnitude of the $(LL)(RR)$ contributions to $B_s$ mixing, as it could easily become of the same order of the $(LL)^2$.

For heavier first generations masses, as in Benchmark 2, we found that in order to avoid tachyons while keeping at least one stop light, it was necessary to use large gluino masses. This spoilt the solution of the flavour tension, unless considerably large $\ord{1}$ parameters were used in the mixing. Although this scenario preserved better the relations between the $(LL)^2$ SUSY contributions to $\Delta F=2$ observables, we found the $(LL)(RR)$ contribution for $B_s$ to be even larger than in the previous benchmark.

The main conclusion for $U(2)^3$ is that it does work as a flavour framework starting at a high scale, preserving most of its virtues without any critical assumptions. From the perspective of solving the flavour tension, this situation changes as the masses of the first two generations are pushed beyond 3~TeV, as the tachyon bound requires heavier gluinos, which in turn spoil the solution of the tension, and give the $(LL)(RR)$ operators further importance.

Finally, we also considered a deviation from minimal $U(2)^3$ breaking, motivated by the need to reproduce neutrino oscillation data. This deviation could induce large contributions to observables in the $K$ sector, spoiling again the correlations. We found that, for both Benchmarks, as long as the deviation was kept of order $\sim\epsilon^2$, the new contributions could be generally considered negligible.

\section*{Acknowledgments}
We would like to thank Gino Isidori for motivating this work and for useful discussions and comments on the draft. G.B.~would like to thank CERN for the hospitality during his visit. J.J.P.~acknowledges partial support from the grants Generalitat Valenciana VALi+d, Spanish MINECO FPA 2011-23596 and the Generalitat Valenciana PROMETEO - 2008/004.

\appendix

\section{$U(2)^3$ Framework}

In the $U(3)^3$ framework of~\cite{Barbieri:2011ci}, the Yukawa matrices were constructed through the addition of three spurions $\Delta Y_u$, $\Delta Y_d$ and $V$, transforming adequately under the flavour symmetries. If the Superpotential is written in the following convention:
\begin{equation}
 W_q=Q_L\,Y_u\,u_R^c\,H_u - Q_L\,Y_d\,d_R^c\,H_d~,
\end{equation}
then the Yukawas acquire the following structure:
\begin{align}
Y_u= y_t \left(\begin{array}{c:c}
 \Delta Y_u & x_t\,V \\ \hdashline
 0 & 1
\end{array}\right), & &
Y_d= y_b \left(\begin{array}{c:c}
 \Delta Y_d & x_b\,V \\ \hdashline
 0 & 1
\end{array}\right)~,
\end{align}
where everything above the horizontal dashed line has two rows, and everything to the left of the vertical dashed line has two columns. The flavour symmetries would allow us to parametrize each $\Delta Y_f$ spurion in terms of its eigenvalues $\lambda_{f_1}$, $\lambda_{f_2}$, and a complex mixing parameter $s_f\,e^{i\alpha_f}$. The $V$ spurion would be described by a suppression parameter $\epsilon$ and complex couplings of $\ord{1}$, $x_f\,e^{i\phi_f}$, with the important fact of having only three independent phases in total. In such a basis, a fit to the CKM matrix performed in~\cite{Barbieri:2011ci} yielded:
\begin{align}
 s_u&=0.095\pm0.008~, & s_d&=-0.22\pm0.01~, \nonumber \\
 s&=0.0411\pm0.0005~, & \cos(\alpha_u-\alpha_d)&=-0.13\pm0.2~,
\end{align}
where $s\propto\epsilon$. Thus, we can choose $\epsilon=\lambda_{\rm CKM}^2$.

Similarly, the sfermion soft masses acquire their structure through the spurions. In this case, the convention for the soft masses shall be:
\begin{equation}
 \mathcal{L}^{\rm soft}_q=\tilde Q^*\,m^2_{\tilde Q}\,\tilde Q^T + \tilde d_R^{c\dagger}\,m^2_{\tilde d}\,\tilde d_R^c
+ \tilde u_R^{c\dagger}\,m^2_{\tilde u}\,\tilde u_R^c~.
\end{equation}

In the unbroken limit, all mass matrices shall have the following structure:
\begin{equation}
m^2_{\tilde f} = \left(\begin{array}{ccc}
m_{f_h}^2 & 0 & 0 \\ 
0 & m_{f_h}^2 & 0 \\ 
0 & 0 & m_{f_l}^2
\end{array}\right)~,
\end{equation}
and we shall assume $m_{f_l}^2<<m^2_{f_h}$. Once we introduce the spurions, the structures of the squark masses becomes:
\begin{eqnarray}
\label{mq.spur}
\frac{m^2_{\tilde Q}}{m^2_{Q_h}} &=&  I+
\left(\begin{array}{c:c} 
c_{Qv}\,V^* V^T   +  c_{Qu}  \Delta Y_u^* \Delta Y_u^{T} 
+  c_{Qd}  \Delta Y_d^* \Delta Y_d^{T} 
&   x_{Q}\, e^{-i\phi_Q} V^* \\ \hdashline
    x_{Q}\, e^{i\phi_Q} V^T   &   -\rho_Q 
\end{array}\right)~,   \\
\frac{m^2_{\tilde d}}{m_{d_h}^2} &=&  I+
\left(\begin{array}{c:c}
c_{dd}\,\Delta Y_d^\dagger \Delta Y_d
&  x_{d}\, e^{-i\phi_d} \Delta Y_d^\dagger V \\ \hdashline
   x_{d}\, e^{i\phi_d} V^\dagger \Delta Y_d  & -\rho_u
\end{array}\right)~,  \\
\frac{m^2_{\tilde u}}{m_{u_h}^2} &=& I+
\left(\begin{array}{c:c} 
c_{uu}\,\Delta Y_u^\dagger \Delta Y_u
&    x_{u}\, e^{-i\phi_u} \Delta Y_u^\dagger V \\ \hdashline
     x_{u}\, e^{i\phi_u} V^\dagger \Delta Y_u   & -\rho_d 
\end{array}\right)~,
\end{eqnarray}
where $\rho_f=(m^2_{f_h}-m^2_{f_l})/m^2_{f_h}$, and all $c_i$ and $x_i$ parameters are real, of $\ord{1}$.

When the Yukawas are diagonalized, the soft matrices are rotated. We are interested in these matrices in the basis where $Y_d$ is diagonal. Such change of basis involves a rotation in the $(2-3)$ block, followed by a further rotation in the $(1-2)$ block. For transparency, we shall write the structure of the soft masses after the first rotation, to leading order in $\epsilon$:
\begin{eqnarray}
\left(\frac{m^2_{\tilde Q}}{m^2_{Q_h}}\right)_{R_{23}} &=&  I-
\left(\begin{array}{ccc} 
0 & 0 & 0 \\
0 & x_{22}\,\epsilon^2 & -x_L\,\epsilon\,e^{i\gamma_L} \\
0 & -x_L\,\epsilon\,e^{-i\gamma_L} & \rho_Q-x_{33}\epsilon^2
\end{array}\right)~,   \\
\label{md:LLGUT}
\left(\frac{m^2_{\tilde d}}{m_{d_h}^2}\right)_{R_{23}} &=&  I-
\left(\begin{array}{ccc}
0 & 0 & 0 \\
0 & 0 & -x_D\,\lambda_{d_2}\,\epsilon\,e^{-i\gamma_D} \\
0 & -x_D\,\lambda_{d_2}\,\epsilon\,e^{i\gamma_D} & \rho_d
\end{array}\right)~,  \\
\label{muLL:GUT}
\left(\frac{m^2_{\tilde u}}{m_{u_h}^2}\right)_{R_{23}} &=& I-
\left(\begin{array}{ccc} 
0 & 0 & 0 \\
0 & 0 & -x_U\,\lambda_{u_2}\,\epsilon\,e^{-i\gamma_U} \\
0 & -x_U\,\lambda_{u_2}\,\epsilon\,e^{i\gamma_U} & \rho_u
\end{array}\right)~.
\end{eqnarray}
Again, $x_i$ are real parameters of $\ord{1}$. These shall be the parameters relevant for phenomenology. In fact, the $\gamma_L$ phase can be identified directly with that appearing in~\cite{Barbieri:2011ci}. Notice that the off-diagonals in $m^2_{\tilde d}$ and $m^2_{\tilde u}$ are suppressed by the second generation quark masses.

Finally, we can apply the rotation in the $(1-2)$ sector, including any further rephasings:
{\small 
\begin{equation}
\label{mqLL:GUT}
\left(\frac{m^2_{\tilde Q}}{m^2_{Q_h}}\right)_{Y_d} =  \left(\begin{array}{ccc}
c_d\,e^{-i(\delta-\alpha_u)} & s_d\,e^{-i(\delta+\alpha_d-\alpha_u)} & 0 \\
-s_d\,e^{i\alpha_d} & c_d & 0 \\
0 & 0 & 1
\end{array}\right)\cdot
\left(\frac{m^2_{\tilde Q}}{m^2_{Q_h}}\right)_{R_{23}}\cdot
\left(\begin{array}{ccc}
c_d\,e^{i(\delta-\alpha_u)} & -s_d\,e^{-i\alpha_d} & 0 \\
s_d\,e^{i(\delta+\alpha_d-\alpha_u)} & c_d & 0 \\
0 & 0 & 1
\end{array}\right), 
\end{equation}}
with a negligible modification of $m^2_{\tilde u}$ and $m^2_{\tilde d}$.

The trilinear couplings follow a structure similar to that of the Yukawas. Their leading structure in the SCKM basis is:
\begin{equation}
\label{trilinear:GUT}
(A_f)_{Y_f} = \left(\begin{array}{ccc}
 a_1 f_1 & 0 & a_2 s_f\,e^{i\alpha_f} \epsilon \\
0 & a_1 f_2 & a_2 c_f \epsilon \\
0 & 0 & a_3
\end{array}\right)y_f A_0~.
\end{equation}


\begin{thebibliography}{99}

\bibitem{D'Ambrosio:2002ex}
  G.~D'Ambrosio, G.~F.~Giudice, G.~Isidori and A.~Strumia,
  Nucl.\ Phys.\ B {\bf 645} (2002) 155
  [hep-ph/0207036].

\bibitem{Dine:1993np}
  M.~Dine, R.~G.~Leigh and A.~Kagan,
  Phys.\ Rev.\ D {\bf 48} (1993) 4269
  [hep-ph/9304299].

\bibitem{Dimopoulos:1995mi}
  S.~Dimopoulos and G.~F.~Giudice,
  Phys.\ Lett.\ B {\bf 357} (1995) 573
  [hep-ph/9507282].

\bibitem{Cohen:1996vb}
  A.~G.~Cohen, D.~B.~Kaplan and A.~E.~Nelson,
  Phys.\ Lett.\ B {\bf 388} (1996) 588
  [hep-ph/9607394].

\bibitem{Giudice:2008uk}
  G.~F.~Giudice, M.~Nardecchia and A.~Romanino,
  Nucl.\ Phys.\ B {\bf 813} (2009) 156
  [arXiv:0812.3610 [hep-ph]].

\bibitem{Barbieri:2010pd}
  R.~Barbieri, E.~Bertuzzo, M.~Farina, P.~Lodone and D.~Pappadopulo,
  JHEP {\bf 1008} (2010) 024
  [arXiv:1004.2256 [hep-ph]].

\bibitem{Badziak:2012rf}
  M.~Badziak, E.~Dudas, M.~Olechowski and S.~Pokorski,
  JHEP {\bf 1207} (2012) 155
  [arXiv:1205.1675 [hep-ph]].

\bibitem{Lunghi:2008aa} E.~Lunghi and A.~Soni, 
  Phys.\ Lett.\ B {\bf 666} (2008) 162
  [arXiv:0803.4340 [hep-ph]].

\bibitem{Buras:2008nn} A.~J.~Buras and D.~Guadagnoli, 
  Phys.\ Rev.\ D {\bf 78} (2008) 033005
  [arXiv:0805.3887 [hep-ph]].

\bibitem{Altmannshofer:2009ne} W.~Altmannshofer, A.~J.~Buras, S.~Gori, P.~Paradisi and D.~M.~Straub, 
  Nucl.\ Phys.\ B {\bf 830} (2010) 17
  [arXiv:0909.1333 [hep-ph]].

\bibitem{Buras:2009pj} A.~J.~Buras and D.~Guadagnoli, 
  Phys.\ Rev.\ D {\bf 79} (2009) 053010
  [arXiv:0901.2056 [hep-ph]].

\bibitem{Barbieri:2011ci}
  R.~Barbieri, G.~Isidori, J.~Jones-Perez, P.~Lodone and D.~M.~Straub,
  Eur.\ Phys.\ J.\ C {\bf 71} (2011) 1725
  [arXiv:1105.2296 [hep-ph]].

\bibitem{Barbieri:2011fc}
  R.~Barbieri, P.~Campli, G.~Isidori, F.~Sala and D.~M.~Straub,
  Eur.\ Phys.\ J.\ C {\bf 71} (2011) 1812
  [arXiv:1108.5125 [hep-ph]].

\bibitem{Barbieri:2012uh}
  R.~Barbieri, D.~Buttazzo, F.~Sala and D.~M.~Straub,
  JHEP {\bf 1207} (2012) 181
  [arXiv:1203.4218 [hep-ph]].

\bibitem{Buras:2012sd}
  A.~J.~Buras and J.~Girrbach,
  arXiv:1206.3878 [hep-ph].

\bibitem{Blankenburg:2012nx}
  G.~Blankenburg, G.~Isidori and J.~Jones-Perez,
  Eur.\ Phys.\ J.\ C {\bf 72} (2012) 2126
  [arXiv:1204.0688 [hep-ph]].

\bibitem{Barbieri:2012bh}
  R.~Barbieri, D.~Buttazzo, F.~Sala and D.~M.~Straub,
  JHEP {\bf 1210} (2012) 040
  [arXiv:1206.1327 [hep-ph]].

\bibitem{Feldmann:2008ja}
  T.~Feldmann and T.~Mannel,
  Phys.\ Rev.\ Lett.\  {\bf 100} (2008) 171601
  [arXiv:0801.1802 [hep-ph]].

\bibitem{Kagan:2009bn}
  A.~L.~Kagan, G.~Perez, T.~Volansky and J.~Zupan,
  Phys.\ Rev.\ D {\bf 80} (2009) 076002
  [arXiv:0903.1794 [hep-ph]].

\bibitem{Xing:2007fb}
  Z.~-z.~Xing, H.~Zhang and S.~Zhou,
  Phys.\ Rev.\ D {\bf 77} (2008) 113016
  [arXiv:0712.1419 [hep-ph]].

\bibitem{Flavour:CKM}
  J.~C.~Hardy and I.~S.~Towner,
  Phys.\ Rev.\ C {\bf 79} (2009) 055502
  [arXiv:0812.1202 [nucl-ex]]~;
  M.~Antonelli, V.~Cirigliano, G.~Isidori, F.~Mescia, M.~Moulson, H.~Neufeld, E.~Passemar and M.~Palutan {\it et al.},
  Eur.\ Phys.\ J.\ C {\bf 69} (2010) 399
  [arXiv:1005.2323 [hep-ph]]~;
  A.~Lenz, U.~Nierste, J.~Charles, S.~Descotes-Genon, A.~Jantsch, C.~Kaufhold, H.~Lacker and S.~Monteil {\it et al.},
  Phys.\ Rev.\ D {\bf 83} (2011) 036004
  [arXiv:1008.1593 [hep-ph]]~;
  R.~V.~Kowalewski [BaBar Collaboration],
  PoS BEAUTY {\bf 2011} (2011) 030.

\bibitem{Martin:1993zk}
  S.~P.~Martin and M.~T.~Vaughn,
  Phys.\ Rev.\ D {\bf 50} (1994) 2282
   [Erratum-ibid.\ D {\bf 78} (2008) 039903]
  [hep-ph/9311340].

\bibitem{ArkaniHamed:1997ab}
  N.~Arkani-Hamed and H.~Murayama,
  Phys.\ Rev.\ D {\bf 56} (1997) 6733
  [hep-ph/9703259].

\bibitem{Tamarit:2012ie}
  C.~Tamarit,
  arXiv:1204.2292 [hep-ph].

\bibitem{:2012gu}
  S.~Chatrchyan {\it et al.}  [CMS Collaboration],
  Phys.\ Lett.\ B {\bf 716} (2012) 30
  [arXiv:1207.7235 [hep-ex]].

\bibitem{:2012gk}
  G.~Aad {\it et al.}  [ATLAS Collaboration],
  Phys.\ Lett.\ B {\bf 716} (2012) 1
  [arXiv:1207.7214 [hep-ex]].

\bibitem{Heinemeyer:1998yj}
  S.~Heinemeyer, W.~Hollik and G.~Weiglein,
  Comput.\ Phys.\ Commun.\  {\bf 124} (2000) 76
  [hep-ph/9812320].

\bibitem{Heinemeyer:1998np}
  S.~Heinemeyer, W.~Hollik and G.~Weiglein,
  Eur.\ Phys.\ J.\ C {\bf 9} (1999) 343
  [hep-ph/9812472].

\bibitem{Degrassi:2002fi}
  G.~Degrassi, S.~Heinemeyer, W.~Hollik, P.~Slavich and G.~Weiglein,
  Eur.\ Phys.\ J.\ C {\bf 28} (2003) 133
  [hep-ph/0212020].

\bibitem{Frank:2006yh}
  M.~Frank, T.~Hahn, S.~Heinemeyer, W.~Hollik, H.~Rzehak and G.~Weiglein,
  JHEP {\bf 0702} (2007) 047
  [hep-ph/0611326].

\bibitem{Buchalla:1995vs}
  G.~Buchalla, A.~J.~Buras and M.~E.~Lautenbacher,
  Rev.\ Mod.\ Phys.\  {\bf 68} (1996) 1125
  [hep-ph/9512380].

\bibitem{Straub:2011fs}
  D.~M.~Straub,
  PoS EPS {\bf -HEP2011} (2011) 146
  [arXiv:1110.6391 [hep-ph]].

\bibitem{Papucci:2011wy}
  M.~Papucci, J.~T.~Ruderman and A.~Weiler,
  JHEP {\bf 1209} (2012) 035
  [arXiv:1110.6926 [hep-ph]].

\bibitem{Baer:2012vr} 
H.~Baer, V.~Barger, P.~Huang and X.~Tata,
  JHEP {\bf 1205}, 109 (2012)
  [arXiv:1203.5539 [hep-ph]]~;
  H.~Baer, V.~Barger, P.~Huang, A.~Mustafayev and X.~Tata,
  [arXiv:1207.3343 [hep-ph]].

\bibitem{Paradisi:2008qh}
  P.~Paradisi, M.~Ratz, R.~Schieren and C.~Simonetto,
  Phys.\ Lett.\ B {\bf 668} (2008) 202
  [arXiv:0805.3989 [hep-ph]].

\bibitem{Hagelin:1992tc}
  J.~S.~Hagelin, S.~Kelley and T.~Tanaka,
  Nucl.\ Phys.\ B {\bf 415} (1994) 293.

\bibitem{Donoghue:1983mx}
  J.~F.~Donoghue, H.~P.~Nilles and D.~Wyler,
  Phys.\ Lett.\ B {\bf 128} (1983) 55.

\bibitem{Gabbiani:1996hi}
  F.~Gabbiani, E.~Gabrielli, A.~Masiero and L.~Silvestrini,
  Nucl.\ Phys.\ B {\bf 477} (1996) 321
  [hep-ph/9604387].

\bibitem{Ciuchini:1998ix}
  M.~Ciuchini, V.~Lubicz, L.~Conti, A.~Vladikas, A.~Donini, E.~Franco, G.~Martinelli and I.~Scimemi {\it et al.},
  JHEP {\bf 9810} (1998) 008
  [hep-ph/9808328].

\bibitem{Raz:2002zx}
  G.~Raz,
  Phys.\ Rev.\ D {\bf 66} (2002) 037701
  [hep-ph/0205310].

\bibitem{Mescia:2012fg}
  F.~Mescia and J.~Virto,
  Phys.\ Rev.\ D {\bf 86} (2012) 095004
  [arXiv:1208.0534 [hep-ph]].

\bibitem{Brod:2011ty}
  J.~Brod and M.~Gorbahn,
  Phys.\ Rev.\ Lett.\  {\bf 108} (2012) 121801
  [arXiv:1108.2036 [hep-ph]].


\end{thebibliography}
\end{document}